\documentclass[twocolumn,floatfix,nofootinbib,british,balancelastpage]{revtex4-1}
\pdfoutput=1
\usepackage{hepunits}
\usepackage{slashed}
\usepackage{multirow}

\usepackage[pdftex]{graphicx}
\usepackage{rotate}
\usepackage{rotating}
\usepackage{relsize}
\usepackage{lineno}
\usepackage{slashed}
\usepackage{balance}
\usepackage{url}
\usepackage{amsmath,mathrsfs}
\usepackage{amssymb}
\usepackage{longtable}
\usepackage{multirow}
\usepackage{tabularx}
\usepackage{units}
\usepackage[usenames,dvipsnames]{xcolor}
\usepackage{subfigure}

\newcommand{\comment}[1]{}

\begin{document}

\title{Exploring QCD uncertainties when setting limits on compressed SUSY spectra}

\author{Herbert Dreiner }%
\email{dreiner@th.physik.uni-bonn.de}
\affiliation{Physikalisches
  Institut and Bethe
  Center for Theoretical Physics, University of Bonn, Bonn, Germany}
\author{Michael Kr\"amer}%
\email{mkraemer@physik.rwth-aachen.de}
\affiliation{Institute for
  Theoretical Particle Physics and Cosmology, RWTH Aachen University,
  Aachen, Germany}
\author{Jamie Tattersall}%
\email{jamie@th.physik.uni-bonn.de}
\affiliation{{}Physikalisches
  Institut and Bethe
  Center for Theoretical Physics, University of Bonn, Bonn, Germany}

\begin{abstract}
 If Supersymmetry (SUSY) has a compressed spectrum the current limits
  from the LHC can be drastically reduced. We take possible `worst
  case' scenarios where combinations of the stop, squark and gluino masses
  are degenerate with the mass of the lightest SUSY particle. To accurately derive limits in the
  model, care must be taken when describing QCD radiation and we examine this
  in detail. Lower mass bounds are then produced by
  considering all the 7~TeV hadronic SUSY and monojet searches. The
  evolution of the limits as the mass splitting is varied is also
  presented.
\end{abstract}

\begin{flushright}BONN-TH-2012-032\end{flushright}
\maketitle

\section{Introduction}

Arguably the best motivated theory for beyond the standard model (BSM) physics is Supersymmetry (SUSY) \cite{Martin:1997ns,Drees:2004jm}. Attractive features of the model include that it leads to a unification of the fundamental couplings, provides the unique way to extend the space-time symmetry and if R-parity is conserved, potentially gives a dark matter candidate with the correct characteristics. However, most relevant for the phenomenology of the Large Hadron Collider (LHC) is that the theory offers a solution to the hierarchy problem.

To solve the hierarchy problem and still remain a natural theory, the mass scale of SUSY must be of the order of the TeV--scale, or lower. Therefore, we can expect that if SUSY exists, it should be probed at the LHC. Unfortunately, whilst the model has now been extensively searched for \cite{:2012rz,Chatrchyan:2012jx,CMS-PAS-SUS-11-022,CMS-PAS-SUS-12-005,:2012mfa}, no hints of SUSY have yet been found and for equal mass squarks and gluinos the limits are above 1.5~TeV for a model with a massless lightest SUSY particle (LSP) \cite{:2012rz}. If we still believe in SUSY as a solution to the hierarchy problem, this leads to two possibilities, either the SUSY mass scale lies just above the limit so far probed or there is some peculiarity in the model of SUSY that makes it particularly difficult to see at the LHC.

The current standard searches rely on two distinctive phenomenological features in order to separate the signal from the background. First of all, the model must produce hard jets and/or leptons so that events can pass experimental triggers. Secondly, under the assumption that the LSP provides the dark matter candidate, a significant amount of missing transverse momentum recoiling from the visible particles must be seen. Therefore, possible ways to hide SUSY are when one or both of these conditions are missing.

One possibility could be that SUSY does not provide the dark matter candidate and we instead have a theory with R-parity violation \cite{Dreiner:1997uz,Barbier:2004ez,Allanach:2003eb}. In this case it is possible that the signal consists only of jets and no missing energy is present in the event. Due to the large multi-jet QCD background with significant uncertainties, such a signal can be very hard to find \cite{Butterworth:2009qa}. For example, in the case that a gluino decays directly to three jets, the bounds on the gluino can be as low as $M_{\tilde{g}} > 280$~GeV \cite{:2012gwa}. In the case of longer decay chains, it is likely that this bound disappears completely \cite{Ruderman:2012jd}.

A second option is that the SUSY spectrum could be compressed with small mass splittings between the coloured superpartners and the LSP. Compression leads to hidden SUSY at the LHC due to the fact that the visible final state particles will only have energies of the order of the mass splitting between the SUSY states. Even if the parent SUSY state is produced with a large boost, relativistic kinematics dictate that the majority of the momentum is transferred to the heavy SUSY daughter (the LSP). Thus, in the limit of a degenerate spectrum, the `hard' event is completely invisible.

Therefore, to see these events, our only option is to look for particles produced in association with the hard event. One possibility is to use hard coloured initial state radiation (ISR), which recoils against the missing momentum of the LSP. This possibility was first investigated as a possible SUSY search mode at the Tevatron \cite{Gunion:1999jr} but large variations in the ISR prediction from parton showers made the mass reach uncertain. Later, the potential of the LHC to search for such topologies was explored in more detail with the use of a matched, matrix element parton shower prediction \cite{Alwall:2008ve,Alwall:2008va,Izaguirre:2010nj}. These techniques were then used to re-analyse LHC searches in order to understand the limits that can currently be extracted \cite{LeCompte:2011cn,LeCompte:2011fh,Dreiner:2012gx,Bhattacherjee:2012mz}. Similar ideas have also been used to look for SUSY top partners (stops) almost degenerate with the LSP, which are motivated by giving a co-annihilation region to predict the correct dark matter relic density \cite{Ajaib:2011hs,He:2011tp,Drees:2012dd,Yu:2012Stop}. Other methods to search for compressed spectra are for example using a monophoton instead of a monojet \cite{Belanger:2012mk} or looking for soft leptons \cite{Rolbiecki:2012gn}.

In order to set reliable mass limits on compressed spectra, the QCD prediction for the radiated jets must be carefully calculated and any uncertainties evaluated. With current tools the radiation can essentially be predicted via two methods, explicit calculation at the parton level using a matrix element or evolution from a hard scale using a parton shower. The matrix element calculation has the advantage that it is exact to fixed order and includes interference between diagrams. However, one can only add a finite number of radiated particles due to computational constraints. In addition, the emissions must be sufficiently hard and well separated to avoid divergences in the result. In contrast the parton shower is only formally correct in the limit of soft and collinear emissions and is only an approximation as we look at harder radiation.

Therefore we must utilise the power of both methods in order to correctly describe hard QCD radiation whilst also generating large particle multiplicities that `look' like real LHC events. Unfortunately we cannot simply generate parton shower events that already contain hard matrix element QCD events because of the possiblity of double counting radiation. Thus, we need to use an algorithm that ensures the possible phase space for emissions is filled only once.

In this project we use two different matching schemes in order to cross check our predictions. The first is the `MLM' matching scheme  \cite{Mangano:2006rw} that has been implemented into MadGraph 5 \cite{Alwall:2008qv,Alwall:2011uj} and interfaced with the Pythia 6 parton shower \cite{Sjostrand:2006za}. This is tested against the CKKW-L algorithm \cite{Catani:2001cc,Lonnblad:2001iq} that has been implemented in Pythia 8 \cite{Sjostrand:2007gs,Lonnblad:2011xx}. In order to fully understand the uncertainties given by the matched predictions, we vary matching, renormalisation and factorisation scales whilst also changing the parton shower properties. 

To make our results widely applicable we avoid looking at specific SUSY breaking scenarios and instead use simplified models. Our philosphy is that we would like to choose models that as far as possible display the `worst case' scenario for SUSY with a dark matter candidate at the LHC. This idea leads us to four benchmark models with the first chosen to place only a single eigenstate squark degenerate with the LSP and all other particles are removed from the spectrum. The second model increases the cross section by having the first two generations of squarks degenerate with the LSP. In the third model we look at gluino limits by placing the particle degenerate with the LSP while all other particles are removed. The final model places both the first two generations of squarks and the gluino degenerate with the LSP. We then investigate how the limits change as we vary the mass splitting to the LSP.

The paper is organised as follows. We start in Sec.~\ref{sec:ISR} by describing the different ways to predict QCD radiation at the LHC and motivating the choice of matching. We also describe in detail, the procedure chosen to estimate uncertainties in our approach. In Sec.~\ref{sec:Simplied_Model}, we motivate and describe each of the simplified models that we have chosen in detail. All current SUSY hadronic searches and monojet analyses are included to set bounds on our models and these are explained in Sec.~\ref{sec:Searches}. The limits derived on our models by these searches are then presented in Sec.~\ref{sec:Limits} and we also briefly discuss how the searches may be improved. In Sec.~\ref{sec:Conclusion} we conclude.

 \section{ISR and matching procedure} {\label{sec:ISR}}
 

\begin{figure}
  \centering
   \includegraphics[width=0.20\textwidth]{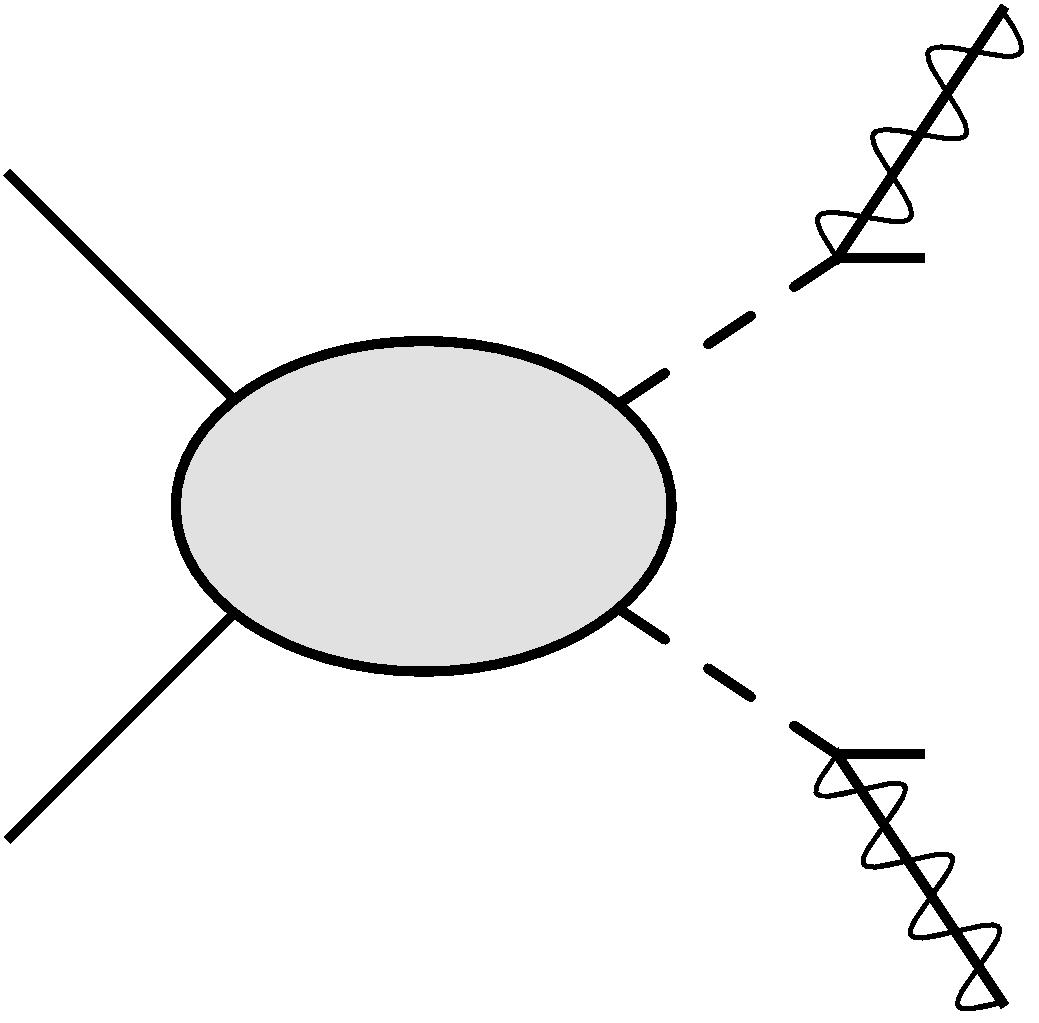}
      \put(-3.9,0.5){$q$}
      \put(-3.9,2.9){$\bar{q}$}
      \put(-1.3,2.5){$\tilde{q}$}
      \put(-1.3,0.8){$\tilde{q}^*$}
      \put(-0.0,3.4){$\tilde{\chi}^0_1$}
      \put(-0.0,0.0){$\tilde{\chi}^0_1$}
      \put(-0.3,2.6){$q$}
      \put(-0.3,0.9){$\bar{q}$}
    \hspace{0.7cm}
   \includegraphics[width=0.22\textwidth]{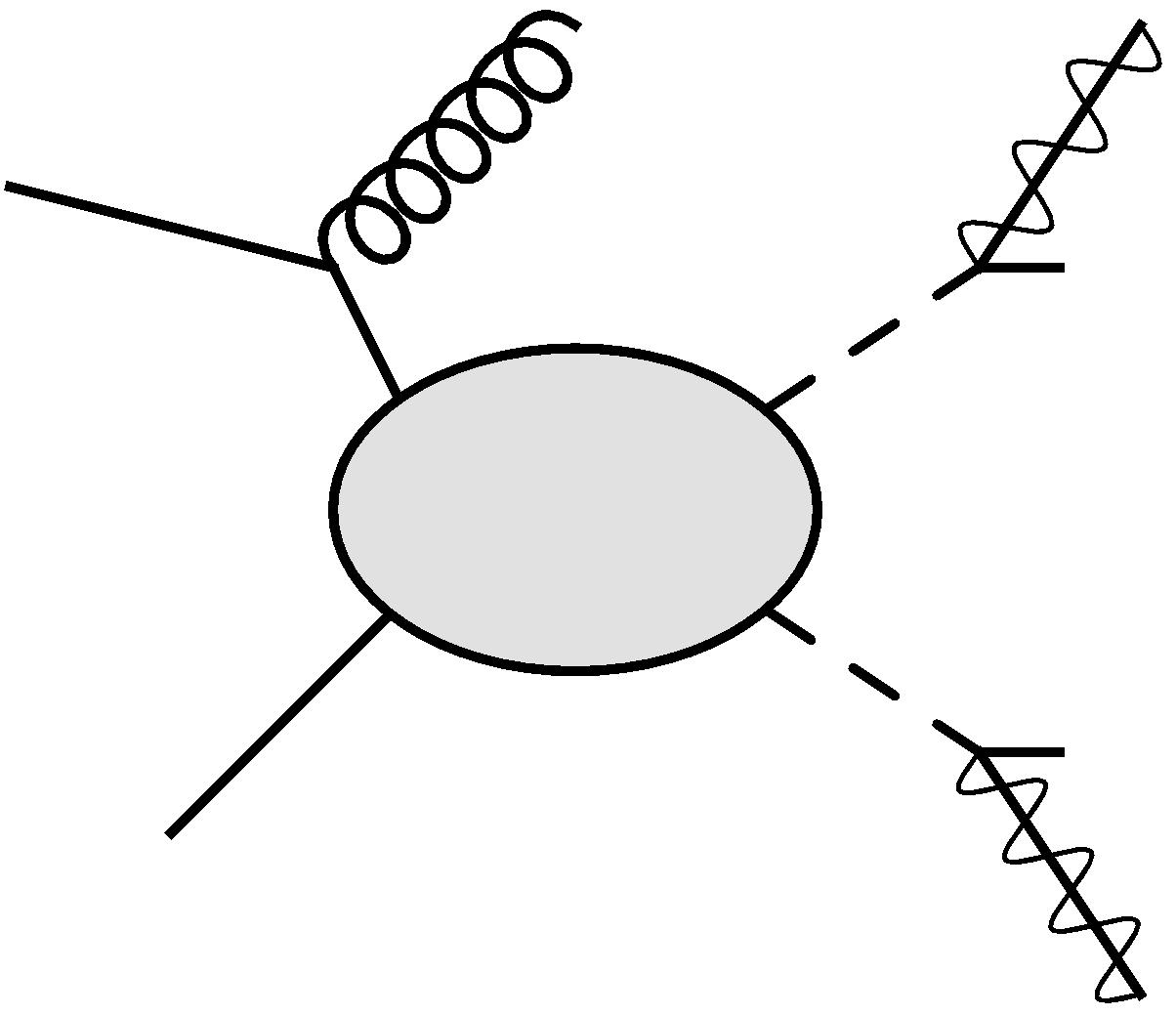}
      \put(-3.7,0.5){$q$}
      \put(-4.2,2.7){$\bar{q}$}
      \put(-1.3,2.3){$\tilde{q}$}
      \put(-1.3,0.8){$\tilde{q}^*$}
      \put(-0.0,3.3){$\tilde{\chi}^0_1$}
      \put(-0.0,0.0){$\tilde{\chi}^0_1$}
      \put(-0.2,2.4){$q$}
      \put(-0.2,0.8){$\bar{q}$}
      \put(-1.9,3.3){$g$}
   \caption{Example event topologies of squark production in compressed spectra with and without an additional radiated jet. For compressed scenarios any jets originating from the final state squark decays are very soft, which is graphically denoted by a short line. \label{fig:ISRDiag}}
\end{figure}
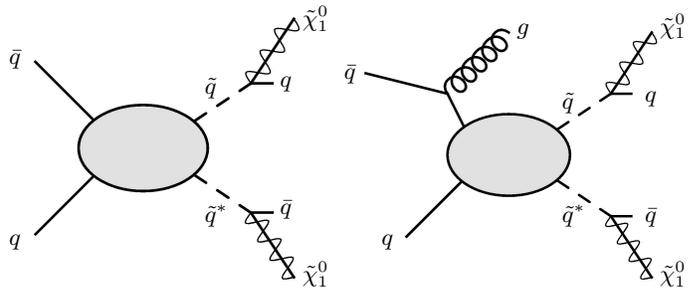

In heavily compressed SUSY spectra, the whole search relies on at least one hard ISR jet to pass any analysis cuts, see Fig.~\ref{fig:ISRDiag} for an example topology. Therefore, the ISR jets must be calculated to as high an accuracy as possible and uncertainties in the prediction must be analysed. However, the hard jet activity is not the only important factor when calculating analysis acceptances. It is vital to also include soft QCD radiation since many cuts can also display a strong dependence on this factor. A particularly clear example is that almost all SUSY search analyses require a minimum angular separation between the MET vector and any jet in the event to significantly reduce the background due to QCD jet mis-measurements. Adding soft QCD radiation significantly increases the number of jets found per event and can thus have a large impact on the number of events that pass these cuts.

\begin{figure*}
  \centering
  \includegraphics[scale=0.11]{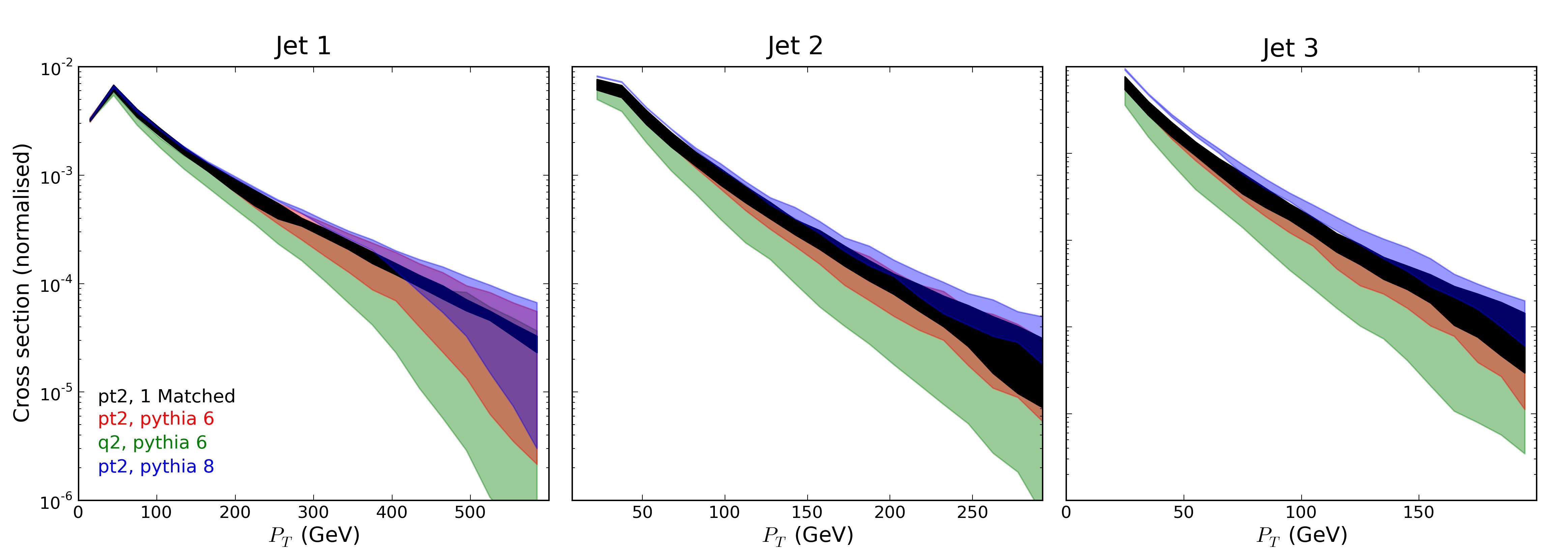}  \hspace{0.8cm}
  \caption{Comparison of the uncertainty associated with 1 jet matched to the parton shower generated events and that of various parton showers without any matrix element emissions. The uncertainty on the parton showers is dominated by varying the starting scale between `wimpy' and `power'. This uncertainty is also included in the matched prediction along with varying the matching scale between 50 and 200~GeV and the factorisation and renormalisation scales simultaneously between $M_T/2$ and $2M_T$. \label{fig:M1vsPS}}
\end{figure*}

In order to predict hard ISR, we have essentially two choices. The first is that we include the extra radiation within the matrix element calculation of the hard event. The big advantage of this approach is that the prediction is exact to the fixed order of the calculation and we include interference effects between different diagrams. We also have a well tested method of estimating the uncertainty in the prediction by varying both the factorisation and renormalisation scales. 

However, the matrix element approach also has its drawbacks. Firstly, the method quickly becomes computationally very expensive as we add more jets. For a SUSY parameter scan, 1 or 2 additional jets can be included, but if we require further radiation, the growth in the number of Feynman diagrams makes this approach prohibitive. In addition, stringent cuts must be placed on the momentum of the extra jets in order not to encounter regions where the perturbative series breaks down due to large logs.

The breakdown in perturbativity can be shown in the cross section for pair production of single eigenstate squarks along with either 1 or 2 jets. We define the usual Durham-$k_{\bot}$ \cite{Brown:1991hx,Catani:1991hj,Catani:1993hr},
	\begin{align}
	     k^2_{\bot} \equiv  \mathrm{min} {\bigg\{} & \mathrm{min} (p^2_{T,i},p^2_{T,j}), \nonumber \\
				&    \mathrm{min} (p^2_{T,i},p^2_{T,j})
					    \frac{(\Delta\eta_{ij})^2+(\Delta\phi_{ij})^2}{D^2} {\bigg\}} 
	\end{align}
with $D=0.4$ to regularise the QCD divergences. The cut can be varied on the additional jets and we show the analytical tree-level cross section in Tab.~\ref{tab:jet_xsec} for the choices, $k^2_{\bot}>25, 50, 100$~GeV.

We show that even if a radiated hard jet of at least 100 GeV is required, the cross section is only decreased by roughly a factor of 4. When two additional radiated jets of 100~GeV are present the cross section is still relatively high and only reduced by a factor of $\sim20$. However, if we reduce the cut to 25~GeV, we see that the one jet cross section is now almost as large as the cross section without a radiated jet. Even the cross section with two additional jets is of the same order. Thus, it is obvious that we can no longer trust the perturbative series at such jet energies.

  \begin{table}[ht!] \renewcommand{\arraystretch}{1.3} \renewcommand{\tabcolsep}{0.3cm}
	 \begin{center}
		 \begin{tabular}{|c|ccc|}\hline
			 Process  & \multicolumn{3}{c|}{Cross section @ 7~TeV (fb)}  \\ 
			$m_{\tilde{q_i}}$ = 500 GeV	      &  $k_T(j) >$ 	& $k_T(j) >$ 	& $k_T(j) >$ \\
							      &   $100$~GeV  	& $50$~GeV 	& $25$~GeV \\  \hline\hline
			 $pp \to \tilde{q}\tilde{q}^*$           & 44.3			&   44.3      		&   44.3  \\
			 $pp \to \tilde{q}\tilde{q}^* \; j$      & 11.5    		&   23.2		&   39.4     \\
			 $pp \to \tilde{q}\tilde{q}^* \; j \; j$ & 1.9    		&   7.5   		&   21.9   \\ \hline
		 \end{tabular}
		\caption{Comparison of analytical tree-level cross-sections (MadGraph) for squark production of a single eigenstate at the LHC with different numbers of radiated jets and different cuts placed on those radiated jets. For simplicity the gluino also has the mass $M_{\tilde{g}}=500$~GeV. \label{tab:jet_xsec} }
	 \end{center}
  \end{table}

  The other approach is to model the QCD radiation with a parton shower. Here the radiation is calculated via a Monte-Carlo program using soft and/or collinear approximations to QCD. The big advantage of this approach is that the large logarithms present are re-summed to give an accurate prediction in the soft QCD regime. Another advantage is that large multiplicity events, that `look' like true LHC collisions are produced. Thus, detector acceptances that can vary heavily with the number of particles in an event can be accurately predicted.

 \begin{figure*}
   \centering
   \includegraphics[scale=0.11]{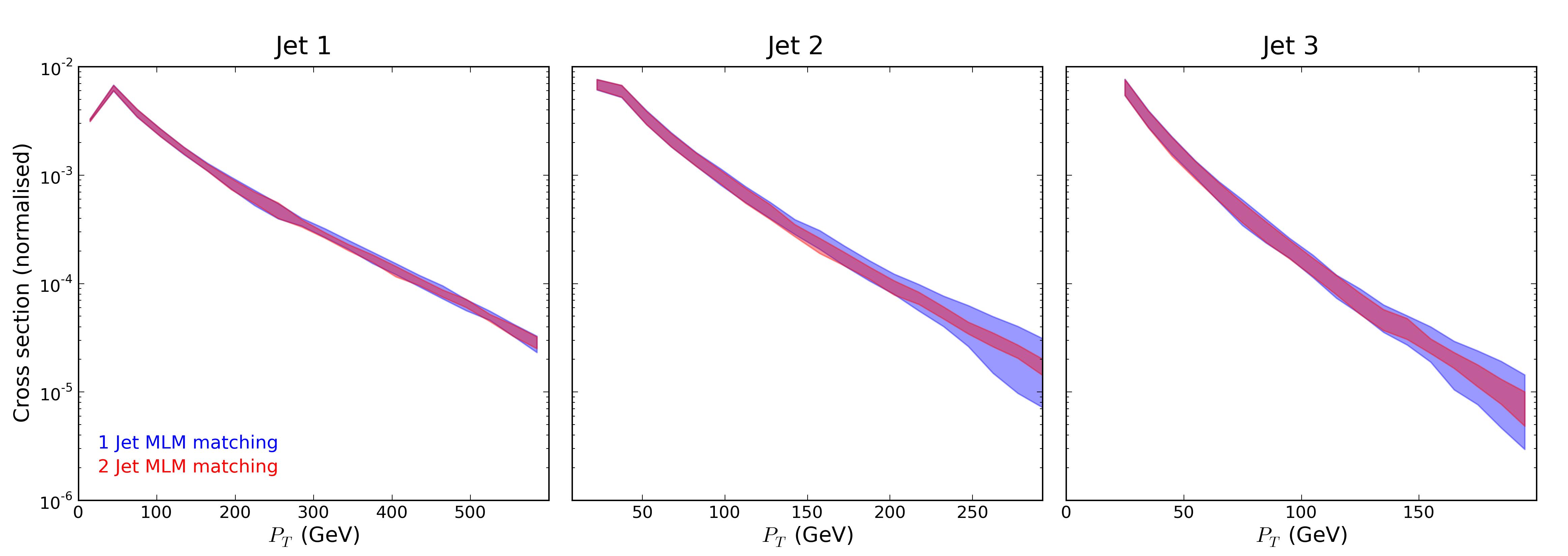}  \hspace{0.8cm}
   \caption{Comparison of the uncertainty associated with 1 jet and 2 jet MLM matching. 
      The uncertainty is found by varying the matching scale between 50 and 200~GeV and the factorisation and renormalisation scales simultaneously between $M_T/2$ and $2M_T$. In addition, the parton showers are varied between the `wimpy' and `power' settings.
 \label{fig:M1vsM2}}
 \end{figure*}

Unfortunately, the approach also suffers from well known deficiencies. Firstly, the parton shower is only an approximation to the matrix element and the prediction degrades as we move away from the soft and collinear limits. In addition, we lose all interference effects. Finally, the parton shower itself is not really predictive for the high energy radiated jets that we will rely on in this study to pass cuts. The reason for this is that the distribution given for these jets is dominated by the scale at which the parton shower is started. Nominally, the starting value should be set to the factorisation scale of the hard process, for example the transverse mass,
  \begin{equation}
	\mu_F = \sqrt{p_T^2 + \hat{m}^2}
  \end{equation}
where $\hat{m}$ is the average mass of the final state particles. A shower that begins at the factorisation scale has until recently been the default for most implementations and has been christened a `wimpy' shower \cite{Plehn:2005cq}.

Although in conflict with the factorisation theorem, more recently it has been shown that a phenomenologically far better approach is to allow the parton shower to fill the full phase space and set the starting scale to the kinematic limit $p_T = \sqrt{s}/2$ \cite{Plehn:2005cq}. This choice has become known as a `power' shower.

In Fig.~\ref{fig:M1vsPS} we show the variation of several versions of the popular Pythia \cite{Sjostrand:2006za,Sjostrand:2007gs} parton shower as we alter the settings between `wimpy' and `power'. We see that for a hardest jet of 600~GeV which is typically the minimum required for our event topologies to pass normal SUSY search cuts, the predicted cross section varies by over three orders of magnitude. Therefore, if our search strategy relies on the presence of ISR to pass the cuts, we cannot rely on the parton shower prediction to tell us the reach of these searches.

Thus, it is obvious that if we want a precision prediction of the ISR we must use the matrix element to calculate hard emissions, while using the parton shower to give us an accurate result in the soft regime and also produce a high multiplicity event. However, this is not as simple as just generating events with additional jets at the matrix element level and then showering these events. If this is done, areas of the phase space will be simultaneously filled by both the matrix element and the parton shower. Therefore, we will double count radiation and predict too hard a spectrum for the additional jets.

The answer to this problem is that we must use an algorithm that successfully matches the matrix element computation to the parton shower generated events whilst making sure that the phase space is only filled once. In addition we would like the matching algorithm to display the following properties. Firstly we want to re-weight different inclusive samples to get a single sample of events. We would also like all distributions to be smooth as we move from the matrix element prediction to the parton shower generated sample. The final result should show only a small dependence on the particular matching scale chosen and settings given in the parton shower. Any remaining difference should be used as a measure of uncertainty in the prediction. Finally, the result should converge to a single prediction as higher multiplicities are added to the matrix element.

\begin{figure*}
  \centering
  \includegraphics[scale=0.11]{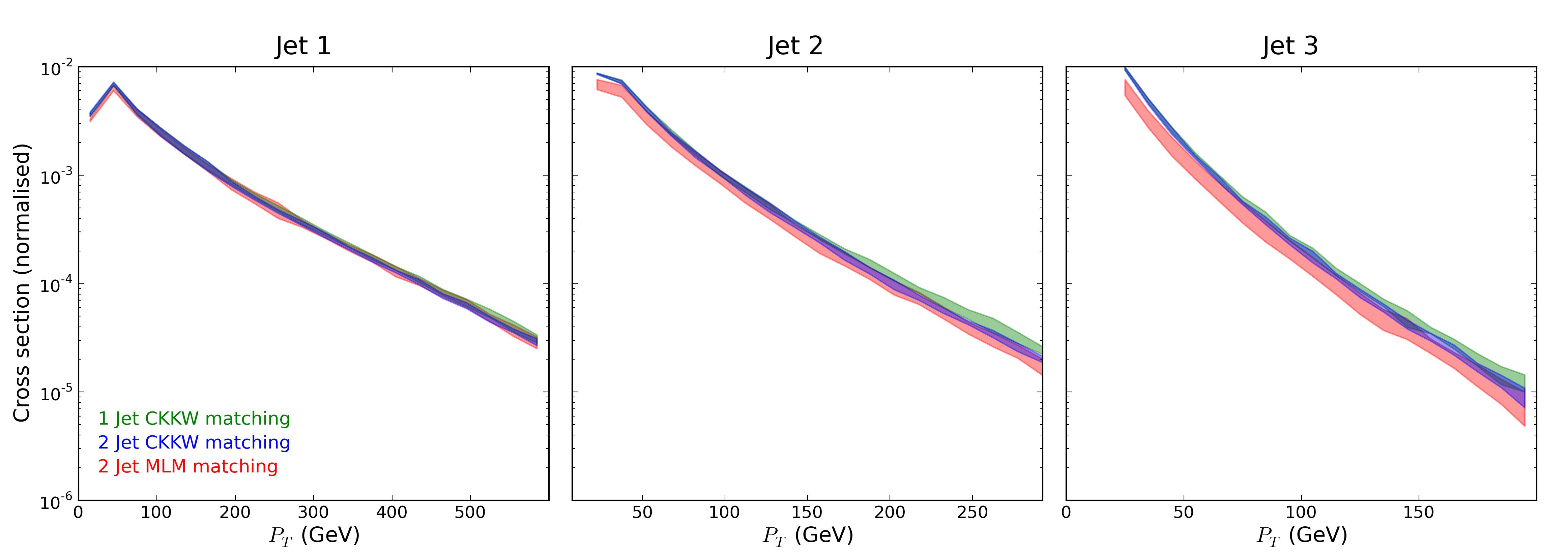}  \hspace{0.8cm}
  \caption{Comparison of the uncertainties between MLM matching in the integrated MadGraph/Pythia 6 algorithm and CKKW-L matching in Pythia 8. The uncertainty is found by varying the matching scale between 50 and 200~GeV and the factorisation and renormalisation scales simultaneously between $M_T/2$ and $2M_T$. In addition, in MLM matching the parton showers are varied between the `wimpy' and `power' settings.\label{fig:MGM1vsP8M2}}
\end{figure*}

In addition to the `normal' double counting issues encountered when matching matrix elements with the parton shower, we must also beware of a second double counting problem within the matrix element calculation of SUSY production itself. The issue is that events with resonant propagators can be double counted and thus must be removed in a consistent way. We use the method detailed in Ref.~\cite{Alwall:2008qv} where events with resonant propagators are removed by hand. This has been shown to work well within the narrow width approximation, $\Gamma/m \ll 1$, that our models always obey. However, it must be stated that interference terms between resonant and non-resonant diagrams are lost. This is in contrast with a proper resonant subtraction procedure \cite{Beenakker:1996ch,Frixione:2008yi} but in our case, the contributions from these terms are small.

In this study two different algorithms are used for matching, to test the predictions and also to provide a consistency check. The first method is MLM matching \cite{Mangano:2006rw} which was implemented within MadGraph \cite{Alwall:2008qv,Alwall:2011uj} and is interfaced with the Pythia 6 shower \cite{Sjostrand:2006za}. In Fig.~\ref{fig:M1vsPS} we show the large reduction in uncertainty that occurs in the prediction of jet radiation when the matching of a matrix element jet to the parton shower is performed. Here, the uncertainty in the matched prediction is estimated by varying the matching scale between 50 and 200~GeV, the parton shower between the `wimpy' and `power' settings and both the factorisation and renomalisation scales simultaneously between $M_T/2$ and $2M_T$ where,
\begin{equation}
      M_T=\sqrt{\sum_i(p_{T_i}^2+M^2_i)}\;.
\end{equation}
Here, $M_i$ and $p_{T_i}$ are the masses and tranverse momentum respectively of the final state particles $i$.

Surprisingly, the reduction in uncertainty not only occurs for the matrix element jet that has been matched but also for jets generated by the parton shower. This is because the uncertainty in the phase space that the second (and third) jet can occupy has been reduced.

Adding a second jet at the matrix element level further reduces the uncertainty in the kinematic distributions, as can be seen in Fig.~\ref{fig:M1vsM2}. The prediction for the hardest jet remains unchanged but we see that the second hardest jet (now given by the matrix element) shows a reduction in uncertainty, as we look at harder emissions. As above, the third jet (produced by the parton shower) prediction is improved due to a reduction in the uncertainty of the phase space available for this emission.

The second method of matching that we use is CKKW-L \cite{Catani:2001cc,Lonnblad:2001iq} that has been implemented in the Pythia 8 Monte Carlo program \cite{Sjostrand:2007gs,Lonnblad:2011xx}. The particular advantage of the CKKW algorithm is that it is sensitive to the internal structure of the matrix element diagrams and thus gives a more consistent treatment of QCD corrections. In addition, the use of Pythia 8 gives a much better description of the underlying event and we were interested to see if this has an effect on our SUSY searches. The disadvantage of the CKKW scheme is that the algorithm is more complicated to implement since it requires an internal interfacing with the parton shower.

In Fig.~\ref{fig:MGM1vsP8M2} we show a comparison between the MLM and CKKW matching predictions. We see that the two methods give consistent results and thus we can be confident that the algorithms are robust. The main difference visible is that the CKKW matched result gives a noticeably harder distribution for softer radiation that is especially visible in the 2nd and 3rd jet. This actually has nothing to do with the matching algorithm but is a result of the more advanced Pythia 8 Monte Carlo program that produces more softer radiation primarily from the underlying event. This extra radiation has been found to be in agreement with LHC data.

\begin{figure*}
  \centering
  \includegraphics[scale=0.11]{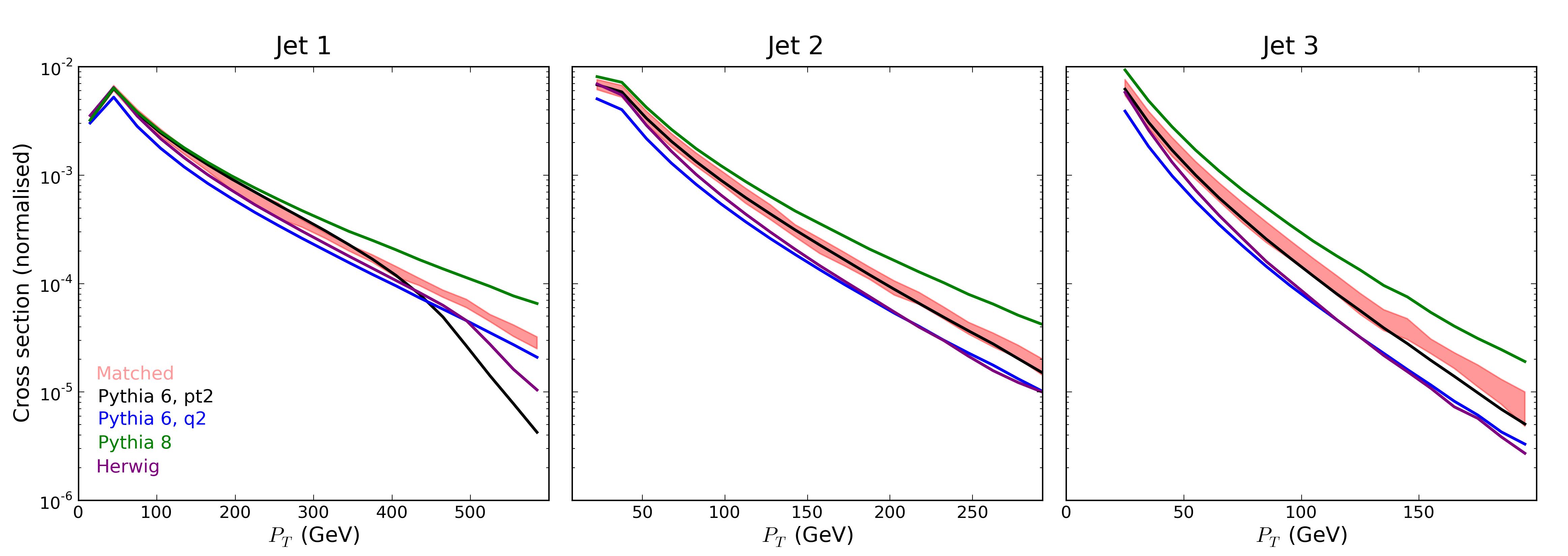}  \hspace{0.8cm}
  \caption{Comparison between 2 jet MLM matching and the default parton shower settings with no matrix element emission for various popular parton showers, Pythia (6+8) and Herwig++.\label{fig:M2vsPSDef}}
\end{figure*}

\begin{figure*}
  \centering
  \includegraphics[scale=0.11]{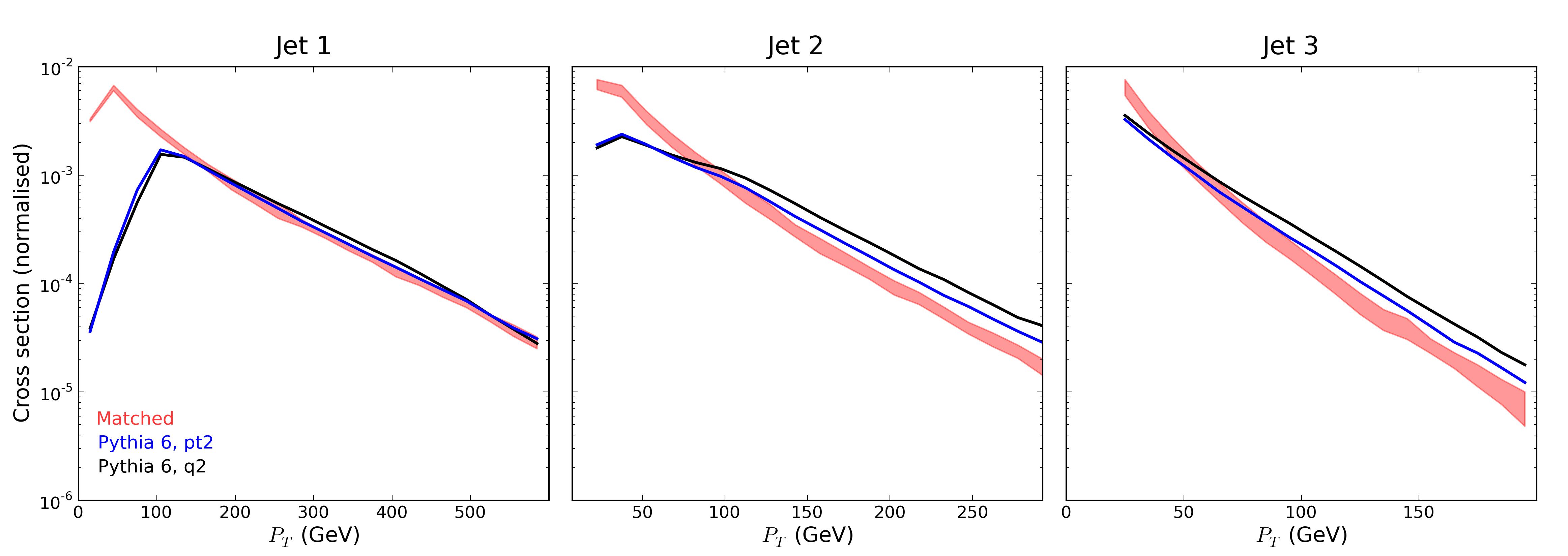}  \hspace{0.8cm}
  \caption{Comparison between 2 jet MLM matching and adding the parton shower to a matrix element that already contains a hard jet (double counting).\label{fig:M2vsPSDoub}}
\end{figure*}

Instead of a comparison between the extremes of parton shower settings available, we can also see how the default behaviour differs from our matched result. In Fig.~\ref{fig:M2vsPSDef} we show the jet distributions for the default settings for different Pythia showers and the Herwig++ shower \cite{Bahr:2008pv}. We see that none of the default choices correctly reproduce the matched result. The Pythia 6 and Herwig++ showers all produce too soft radiation, while the Pythia 8 shower (which is a `power' shower by default) actually gives too hard a spectrum. In total, we see a variation of well over 1 order of magnitude depending on the particular parton shower chosen for jet $p_T>600$~GeV.

A final comparison we would like to make is between our matched result and the distribution that would be found if you simply take matrix element events that already contain extra jets and apply a parton shower. As stated before, this method is inconsistent since the phase space for extra jets can be filled by both the matrix element and parton shower. However, we show the results here to investigate the size of the error induced by this approach. For this comparison we use the relatively soft Pythia 6 default shower in order not to overstate the problem. We find that the choice actually gives a reasonable result for the hardest jet in the event. The reason is that with this choice of `soft' shower, in general, the hardest radiation will come from the matrix element. Therefore, the distribution for the hardest jet is not altered by too much. However, if we look at the second and third jets we see that the distributions have become significantly harder. The reason is that events are now produced where both the matrix element and the the parton shower have given a `hard' jet. Thus we see that we have a significant double counting problem but by only looking at the hardest jet distribution you may have concluded otherwise.

 \section{Simplified Models} {\label{sec:Simplied_Model}}
 \subsection{Motivation}
We would like to set lower mass limits on the R-parity conserving spectrum that are robust and involve the minimum number of assumptions. For this reason we use a range of simplified models to capture the `worst case' for the discovery potential of gluinos or squarks at the LHC.

\begin{figure*} 
  \hspace{1.4cm}
  \begin{center} 
    \includegraphics[scale=0.12]{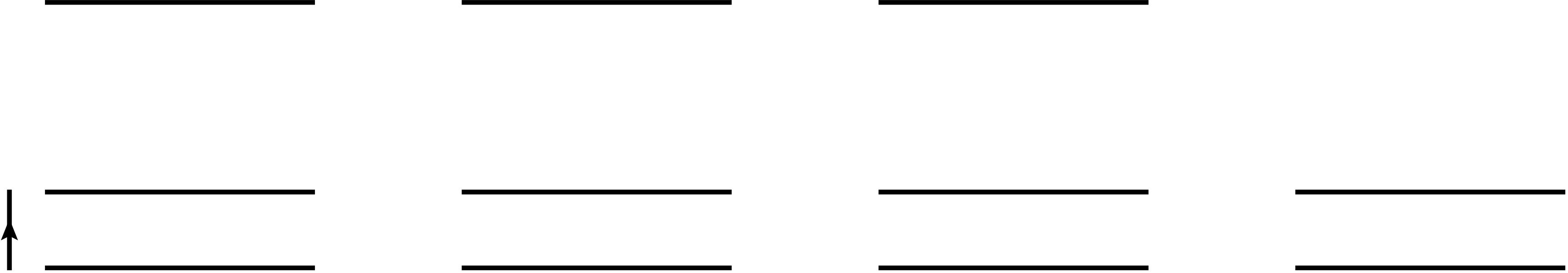}
 	   \put(-14.5,3.3){(a) \bf{Stop (single}}
	  \put(-13.8,2.8){\bf{eigenstate)}}
 	  \put(-10,3.3){(b) \bf{Squark}}
 	    \put(-6.1,3.3){(c) \bf{Gluino}}
 	    \put(-2.8,3.3){(d) \bf{Equal mass}}
 	      \put(-13.5,2.0){$\tilde{g},\; \tilde{q}_{1,2}$}
 	      \put(-9.5,2.0){$\tilde{g}$}
 	      \put(-5.5,2.0){$\tilde{q}_{1,2}$}
 	      \put(-13.5,1.0){$\tilde{t}_1$}
 	      \put(-9.5,1.0){$\tilde{q}_{1,2}$}
 	      \put(-5.5,1.0){$\tilde{g}$}
 	      \put(-1.8,1.0){$\tilde{g}\sim\tilde{q}_{1,2}$}
 	      \put(-13.5,-0.5){$\mathrm{LSP}$}
 	      \put(-9.5,-0.5){$\mathrm{LSP}$}
 	      \put(-5.5,-0.5){$\mathrm{LSP}$}
 	      \put(-1.8,-0.5){$\mathrm{LSP}$}
 	      \put(-14.6,2.4){$\infty$}
 	      \put(-15.35,0.2){$\Delta M$}
  \end{center} 
  \caption{The spectra for the simplified models studied in this paper. For the \textquoteleft Stop' scenario we place the stop (or a single eigenstate squark) quasi-degenerate with the LSP and remove all other particles from the spectrum. In the \textquoteleft Squark' scenario we place the first and second generation squarks quasi-degenerate with the LSP whilst removing all other particles. The `Gluino' scenario has the gluino placed quasi-degenerate with the LSP and all other particles removed. In the `Equal mass' scenario, the first two generations of squarks and the gluino is placed degenerate with the LSP and all other particles are removed. In all models quasi degenerate refers to 1~GeV mass splitting. Larger mass splittings are investigated all the way up to a massless LSP. \label{fig:Spectrum}}
\end{figure*}
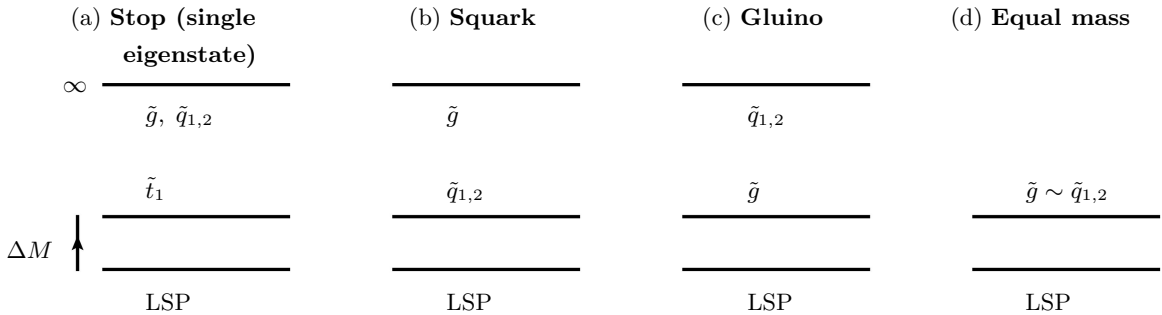

In R-parity conserving SUSY, this worst case behaviour is found by placing the produced particles of interest, quasi-degenerate with the LSP. The degeneracy results in all the momentum carried by the parent particle being passed to the massive LSP whilst 
almost no momentum is given to the Standard Model particle. Consequently, the sparticle decay is invisible to the detector, as all momentum is carried by the invisible LSP. Therefore, the only way that events can be seen at the LHC is via the emission of hard QCD radiation. We study these models from quasi-degeneracy (1~GeV mass splitting) to the kinematic limit where the LSP is massless\footnote{We note that measuring a LSP neutralino mass below 10 GeV at colliders is very difficult
\cite{Dreiner:2009er,Conley:2010jk,Choi:2011vv,Dreiner:2012ex}.} to see how the bounds evolve.

Despite us labeling our scenarios as the `worst case' a few caveats must be added for possible exceptions. Firstly, we assume that all decays are prompt and therefore, although our spectra are compressed, they are not so compressed that they lead to displaced vertices or long lived states. However, we believe that both of these possibilities would only strengthen the bounds on the model. A charged and/or coloured particle traversing the detector produces a very distinctive signal that should make such a model easy to detect \cite{:2012vd,Chatrchyan:2012sp}. In addition, displaced vertices offer another handle with which to discover a model.

Another assumption made is that no other states exist between the produced mother particle and the final state LSP. If other states exist this could potentially lead to longer and more complicated decay chains. In the limit of degeneracy (and assuming prompt decays), these extra states will not change the phenomenology of the model from an LHC perspective because all of the mother's particle momentum will still be transferred to the LSP. However, as the mass splitting is increased, more momentum can be potentially transferred to intermediate states which will result in more soft jets but less missing energy. Therefore placing many intermediate states with small mass splittings may offer a different way to hide SUSY at the LHC \cite{Fan:2012jf}.

\subsection{Models Considered}

The simplified models that we use are all depicted in Fig.~\ref{fig:Spectrum}. We label our first scenario `Stop' and here we take a single eigenstate stop quark quasi-degenerate with the LSP while all other particles are removed from the spectrum. We assume for this simplified model that the stop decays to a single light jet and the LSP. An example of this decay mode in SUSY is the well known loop induced decay \cite{Hikasa:1987db,Boehm:1999tr,Muhlleitner:2011ww,Drees:2012dd},
\begin{equation}
    \tilde{t}_1 \to c \;  \tilde{\chi}^0_1,
\end{equation}
that often dominates in scenarios where the stop is close in mass to the LSP.

As stated above, in the limit of degeneracy, the exact decay mode of the stop quark does not change the phenomenology. However, as we increase the mass splitting, the precise decay mode can become important. An example is that if the final state decay products are $b\ell\nu$ or $bjj$, the process can look very similar to normal top production. In this case, the bounds may be weakened from the ones presented here.

We would also like to state, that although we have labeled the scenario `Stop', it can in fact apply to any scenario where only a single squark eigenstate is light. This is because if the gluino is decoupled, there are no t-channel production processes. Thus the cross section is identical for all squark states and there is no PDF flavour dependence. In fact in this case, the limit is more likely to hold as the mass splitting is increased between the squark and the LSP because the decay,
\begin{equation}
    \tilde{q}_i \to q_i \; \tilde{\chi}^0_1,
\end{equation}
can be expected to dominate.

The second model that we consider, `Squark', can almost be considered a sub-category of the above. Here, instead of a single eigenstate, we now place the first two generations of squarks, quasi-degenerate with the LSP. Thus, the only difference in phenomenology from the first model mentioned is that the cross section is increased by a factor of eight. Hence, we can expect the limit to be significantly more stringent. The model can easily accommodate a different number of squarks by a simple rescaling of the cross-section. After considering the quasi-degeneracy case ($\Delta M = 1$~GeV), we then study the phenomenology as the mass splitting is increased, until we reach the limit of a massless LSP. These models can be in particular motivated if the gluino is given a Dirac mass term, that can produce a heavy gluino ($\sim 5$~TeV) whilst still being technically natural \cite{Kribs:2012gx}.

For our third model, `Gluino', we essentially flip the spectrum compared to the first two. Here we place the gluino quasi-degenerate with the LSP whilst all other particles are removed. In the limit that all squarks are removed from the model, it must be stated that the gluino becomes stable. Therefore, if this model was realised in nature we would see the distinctive signal of so called `R-hadrons' in the detector. However, it is possible that the third generation squarks could be much lighter than the other squarks. These could mediate prompt gluino decay whilst having a negligible impact on the search. Therefore we assume a prompt decaying gluino in this scenario as an interesting limiting case. Once again, after considering the quasi-degenerate case, we investigate increasing the mass splitting.

The final scenario we call, `Equal mass', and here we consider a model where the first two generations of squarks and the gluino are both quasi-degenerate with the LSP but increased mass splitting to the LSP is again considered.

In the last three models mentioned we essentially ignore the third generation from our studies. The rational for this treatment is that in many high scale models these states are generically split from the other squarks by the large Yukawa coupling in the Renormalisation Group Equations (RGEs). However, it is also true that they would have minimal impact on the searches due to their low cross-sections, but removing them produces a more conservative limit.

 \section{Searches} {\label{sec:Searches}}

In order to set the best possible limits on each of our simplified
models we apply all of the 7~TeV ATLAS and CMS hadronic SUSY
searches. The motivation behind exploring all of the searches is that
essentially two different strategies now exist for searching for SUSY
and we would like to understand which is the most effective method for
compressed spectra. The `traditional' method is to search for hard
jets and a significant proportion of missing transverse energy (MET)
which provides the discriminator between signal and background. Both
ATLAS \cite{:2012rz} and CMS \cite{:2012mfa} have searches of this
kind.

A second method is to try and use some kind of topological cut that better separates signal and background. Hence, these cuts may be able to set kinematic cuts a little lower and thus improve signal acceptance. CMS has three different searches of this kind, razor \cite{CMS-PAS-SUS-12-005}, $\alpha_T$ \cite{CMS-PAS-SUS-11-022} and $M_{T2}$ \cite{Chatrchyan:2012jx}.
  
The expected topology from compressed spectra events is expected to be a single hard jet balanced by missing energy from two invisible LSPs, see Fig.~\ref{fig:ISRDiag}. Therefore it is natural to also look at monojet searches at the LHC and examine how these can constrain our models. Both CMS \cite{Chatrchyan:2012me} and ATLAS \cite{ATLAS:2012ky} have such a monojet search and we include both in this study.

All the searches were implemented within the analysis program RIVET \cite{Buckley:2010ar}. In order to better test experimental effects, momentum smearing and mis-measured tails on jets were included \cite{Allanach:2011ej}, but were found to have a negligible impact on search reach. The searches were tested against all the mSUGRA and simplified models presented in each individual analysis. In addition, whenever cut flows and kinematic distributions were presented for individual MSSM points, these were also compared against. The agreement was always found to be within 20\% but was usually much better. We use the Rolke Test \cite{Rolke:2000ij,Rolke:2004mj,Lundberg:2009iu} to derive the 95\% confidence level exclusion for each model and search. The exclusion is derived by using the search region box with most discriminating power. However, if any box contains an under-fluctuation in data, we use the expected limit with no under-fluctuation. This leads to a more conservative bound than the experiments quote but 
allows for a fairer comparison between different analysis strategies. Exceptions and differences between our implementation and the official analyses were most notably found between the different statistical methods used to derive limits and these are described below in the search discussions.

\subsection{Jets and MET searches}

We begin by describing the `vanilla' jet and MET searches that are used by both ATLAS \cite{:2012rz} and CMS \cite{:2012mfa}. The basic idea of both these searches is to use hard reconstructed jets and a significant proportion of MET to discriminate the signal from the background.

The ATLAS baseline selection requires at least two jets in the event with $p_T(j)>60$~GeV and the harder jet having $p_T(j_1)>130$~GeV. In addition, a minimum MET of $E_T^{miss} > 160$~GeV is required (although all the search regions relevant for our study require significantly more) with a minimum angular distance between the jet and MET vectors of $\Delta\phi(jet,E_T^{miss})>0.4$ to protect against QCD mis-measurement. For the search region, a variable is defined to give an approximate measure of the SUSY scale,
\begin{equation}
      m_{eff}=\sum_{N jets}p_T(j) + E_T^{miss}.
\end{equation}
The search is then divided into 11 boxes that require different combinations for the number of jets (2 - 6), $m_{eff}$ (900 - 1900~GeV) and $E_T^{miss}/m_{eff}$ (0.15 - 0.4).

Limits are set on SUSY models by first determining which search box provides the best discriminating power for a particular area of parameter space. Once the box has been determined, the limit is found by using the real data from the experiment. This is a different procedure from the method that we use (explained above) because under-fluctuations in the data can produce a better bound than initially expected. As stated, in order to be able to provide a fairer comparison between different searches, we use the expected limit in the case of under-fluctuations but it must be noted that this produces a more conservative limit.

The CMS search is very similar in philosophy with the most important difference for our models being that the baseline selection now requires 3 jets but these can be slightly softer with $p_T>50$~GeV. For the search, a slightly different variable is used, 
\begin{equation}
    H_T=\sum_{N jets}p_T(j),
\end{equation}
and this time 14 search regions are defined with different combinations of $H_T$ (500 - 1400~GeV) and $E_T^{miss}$ (200 - 600~GeV). To set limits, CMS uses a test statistic that combines all bins but we again only take the single region with most discriminating power. As before, this leads to more conservative limits.

\subsection{Topology based searches}

As mentioned above, in addition to the normal jet and MET searches, CMS also has a range of topology based searches. The idea of these is that an event shape cut is used to better discriminate signal and background.

The first such search that we will mention is the `razor' analysis \cite{CMS-PAS-SUS-12-005}. The initial baseline requires at least two jets with $p_T>50$~GeV and the first step in the analysis is to combine all final state jets into two so-called `megajets' that can be of any size. From the two megajets two new variables are formed, first the longitudinal boost invariant,
\begin{equation}
      M_R \equiv \sqrt{(E_{j1} + E_{j2})^2 - (p_z^{j1}+p_z^{j2})^2)} \;,
\end{equation}
that can be expected to peak at the SUSY mass scale over a falling Standard Model background. Second a variable that acts like the average transverse mass of the megajets,
\begin{equation}
      M_T^R \equiv \sqrt{ \frac{E_T^{miss}(p_T^{j1} + p_T^{j2}) - \vec{E}_T^{miss}\cdot(\vec{p}_T^{j1} + \vec{p}_T^{j2})}{2}} \;,
\end{equation}
where $\vec{E}_T^{miss}$ is the missing transverse momentum ($\vec{p}_T^{miss}$) and the magnitude is $E_T^{miss}$.
The two variables are then put together to give the razor dimensionless ratio,
\begin{equation}
      R \equiv \frac{M^R_T}{M_R} \;.
\end{equation}
For SUSY events with genuine missing energy, $M^R_T$ will approximately be maximal for $M^R_T=M_R$ and thus $R$ will have a maximum of 1. However the distribution will peak at roughly $R=0.5$ in contrast to QCD multi-jet events that will peak at 0.

Limits are set using a complicated variable binned likelihood in $M_R$ (500 - 2000~GeV) and $R^2$ (0.18 - 0.5) that is unfortunately impossible to replicate without the unbinned data. However, a 60 bin data set was made available \cite{CMSRazor:wiki} and to replicate the statistical methods we used in the other searches, we reduced this to 20 bins. With our method we see a noticeable reduction in the search reach but we believe this is the fairest way to compare the razor variable with other search techniques.

\begin{table*}[ht] \renewcommand{\arraystretch}{1.2} \renewcommand{\tabcolsep}{0.3cm}
\begin{center}
\begin{tabular}{|c||c|c|cccc|} \hline 
					& 				& Search Region			& 	 \multicolumn{4}{c|}{Degeneracy Bound (GeV)}  	 \\
 Search 				& $\mathcal{L}$ (fb$^{-1}$) 	& (given in source)		&Stop		&Squark  	& Gluino  	       & Equal \\ \hline\hline
 \underline{Monojet}			&				&				&		&		&		&\\
 \bf{ATLAS*} \cite{ATLAS:2012ky}		& 	$4.7$			&  SR3/SR4			&\bf{230}	&\bf{370} 	& \bf{520} 	& \bf{680} \\ 
 CMS*  \cite{Chatrchyan:2012me}		& 	5.0			&  $E_T^{\mathrm{miss}} > 400$	&$190$ 		& $340$ 	& $480$ 	& $650$  \\ \hline\hline
 \underline{SUSY}			&				&  				&		&		&		&	\\
  ATLAS MET \cite{:2012rz} 		& 	4.7			&	A' med/C med		&- 		&$260$ 		& $450$ 	& $540$  \\ 
   CMS $\alpha_T$ \cite{CMS-PAS-SUS-11-022} & 	$5.0$			&  Optimised $H_T$ bin 		&$190$ 		&$330$ 		& $530$ 	& $600$  \\ 
   CMS MET \cite{:2012mfa}		& 	5.0			&	A2			&- 		&$300$ 		& $460$ 	& $550$  \\ 
   CMS $M_{T2}$ \cite{Chatrchyan:2012jx} & 	4.7			&	A/B			&- 		&- 		& $400$		 & $500$  \\ 
   \bf{CMS Razor} \cite{CMS-PAS-SUS-12-005} &	4.4			&	bHad($6_4+7_4+8_4+9_4$)	&\bf{200} 	&  \bf{350} 	& \bf{530} 	& \bf{610}  \\  \hline
\end{tabular}
\caption{Comparison of the bounds on the mass of SUSY particles for the different searches employed at the LHC. The luminosity of the searches and the most constraining search region are also given (the search region names refer to those given in the original experimental papers). *\textit{The ATLAS and CMS monojet searches only give these bounds for mass differences $<5$~GeV. For larger mass splittings, the bounds become much weaker.}  \label{tab:Limits} }
\end{center}
\end{table*}

A similar topology based search is made using the $\alpha_T$ variable \cite{CMS-PAS-SUS-11-022}. The initial selection requires that at least two jets have $E_T > 100$~GeV and that $H_T > 275$~GeV. If more than two jets are present in the event they are clustered in such a way to minimise the difference in $E_T$ between the resulting pseudo-jets. The discriminating variable is defined as,
\begin{equation}
     \alpha_T \equiv \frac{E_T^{j2}}{\sqrt{H_T^2 - \slashed{H}_T^2}},
\end{equation}
where $E_T^{j2}$ is the scalar sum transverse energy of the softer pseudo-jet and
\begin{equation}
    \slashed{H}_T \equiv \sum_{N jets}\vec{p}_T(j).
\end{equation}
For perfectly measured back to back QCD jets, $\alpha_T=0.5$ but if either of the jets is mis-measured, this will lead to values of $\alpha_T$ less than 0.5. However, in events that contain real missing energy that the jets recoil against, far larger values of $\alpha_T$ can be seen and this provides an effective discriminator. For multi-jet QCD final states, it is possible that large mis-measurements can lead to values slightly above 0.5 so a cut of $\alpha_T>0.55$ is placed to remove this background.

Limits are set with a binned likelihood over 8 signal regions of varying $H_T$ (275 - 875~GeV). We again only use the signal region with most discriminating power which leads to a more conservative limit.

The final topology based search we consider uses the $M_{T2}$ variable \cite{Chatrchyan:2012jx}. The initial baseline requires at least three jets to be reconstructed with $p_T > 40$~GeV and two of these jets must have $p_T > 100$~GeV. In addition a selection of $H_T > 650$~GeV is required. As the discriminating variable, the search uses a simplified version of $M_{T2}$ where two massless pseudo-jets are formed and the LSP is assumed to be massless,
\begin{equation}
      M_{T2} \equiv \sqrt{2 p_T^{j1} p_T^{j2} (1+\cos\phi_{12})}.
\end{equation}
To set limits, 10 bins are defined in the variables $H_T$ (750 - 950~GeV) and $M_{T2}$ (150 - 500~GeV). A multi-bin profile likelihood is then used in the official analysis but we only use the signal region with the most discriminating power which once again sets a conservative limit.

\subsection{Monojet Searches}

In addition to the SUSY searches mentioned previously, we also consider the reach of searches primarily designed for a monojet topology. One of the motivations of these studies has been to look for model independent dark matter when an ISR jet recoils from the pair production of WIMPS. In compressed SUSY, the event signal will be identical and thus we hope that these searches may lead to competitive bounds.

We begin by describing the ATLAS monojet search \cite{ATLAS:2012ky}
which requires at least one jet with $p_T>120$~GeV and also missing
energy $E_T^{miss}>120$~GeV. A veto on events with a third jet,
$p_T>30$~GeV, is in place but a second jet is allowed as long as
$\Delta\phi(\vec{p}_T^{miss},\vec{p}_T^{j2})>0.5$. Four search regions
are defined within symmetrical requirements for the hardest jet $p_T$
and missing energy $E_T^{miss}$ varying between 120 and 500~GeV. To
set limits, only the region with the best exclusion is used. Our
method is almost the same apart from the fact that two of these search
regions contain under-fluctuations and hence we use the expected limit
here.

The CMS monojet search \cite{Chatrchyan:2012me} is very similar in
philosophy and only contains small numerical differences in the cuts
applied. For the initial selection, the jet requirement is softened
slightly with $p_T>110$~GeV but this missing energy is hardened
$E_T^{miss}>200$~GeV. However, for the search regions relevant to our
study, these differences are inconsequential. Again events with a
third jet $p_T>30$~GeV are vetoed and the second jet direction cut is
tightened slightly with
$\Delta\phi(\vec{p}_T^{j1},\vec{p}_T^{j2})<2.5$. Four search regions
are also considered but now only missing energy is used as a single
discriminator with $E_T^{miss}$ varying between 250 and 400~GeV.

 \section{Limits} {\label{sec:Limits}}

\begin{figure*}
  \centering
  \includegraphics[width=0.49\textwidth]{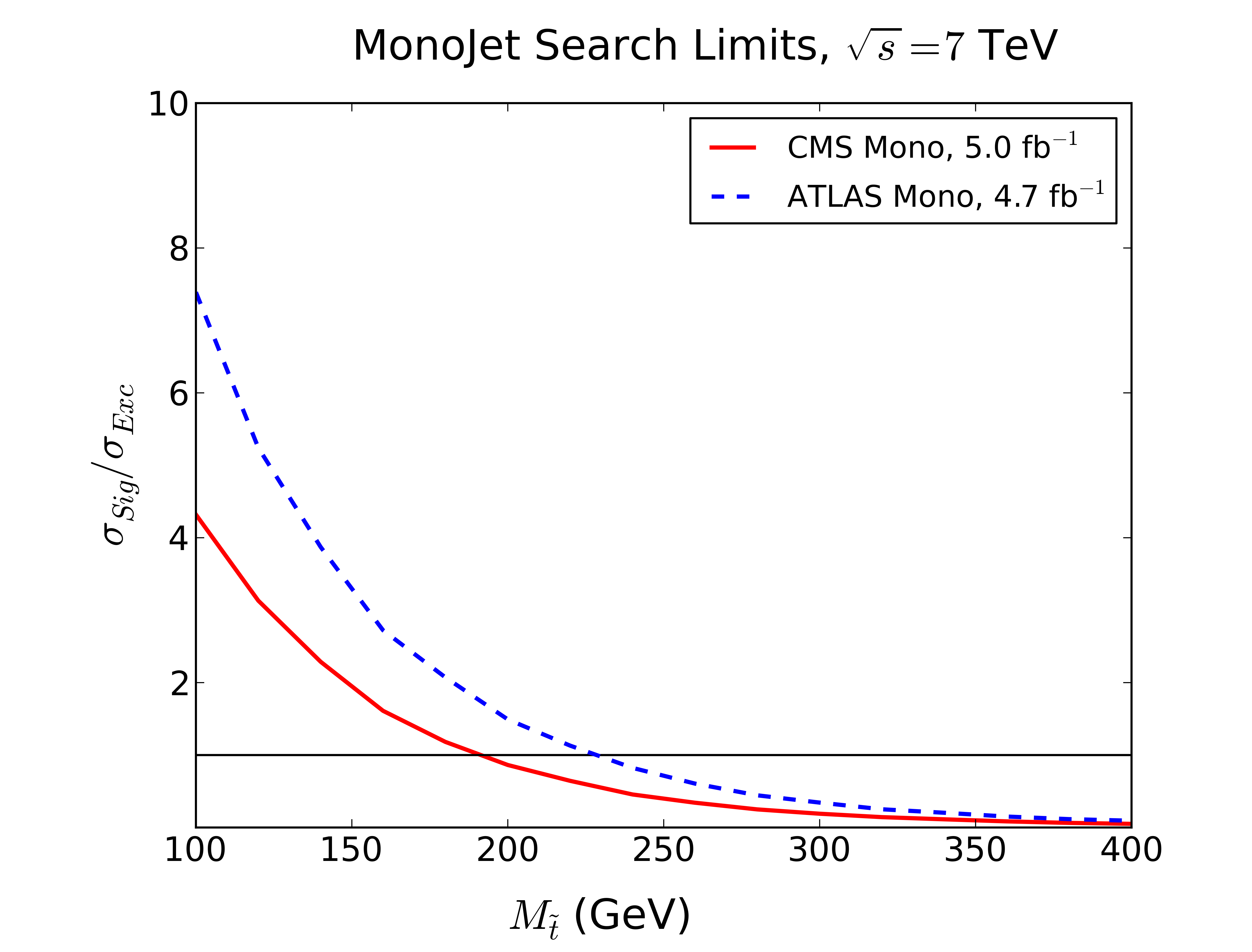}
  \includegraphics[width=0.49\textwidth]{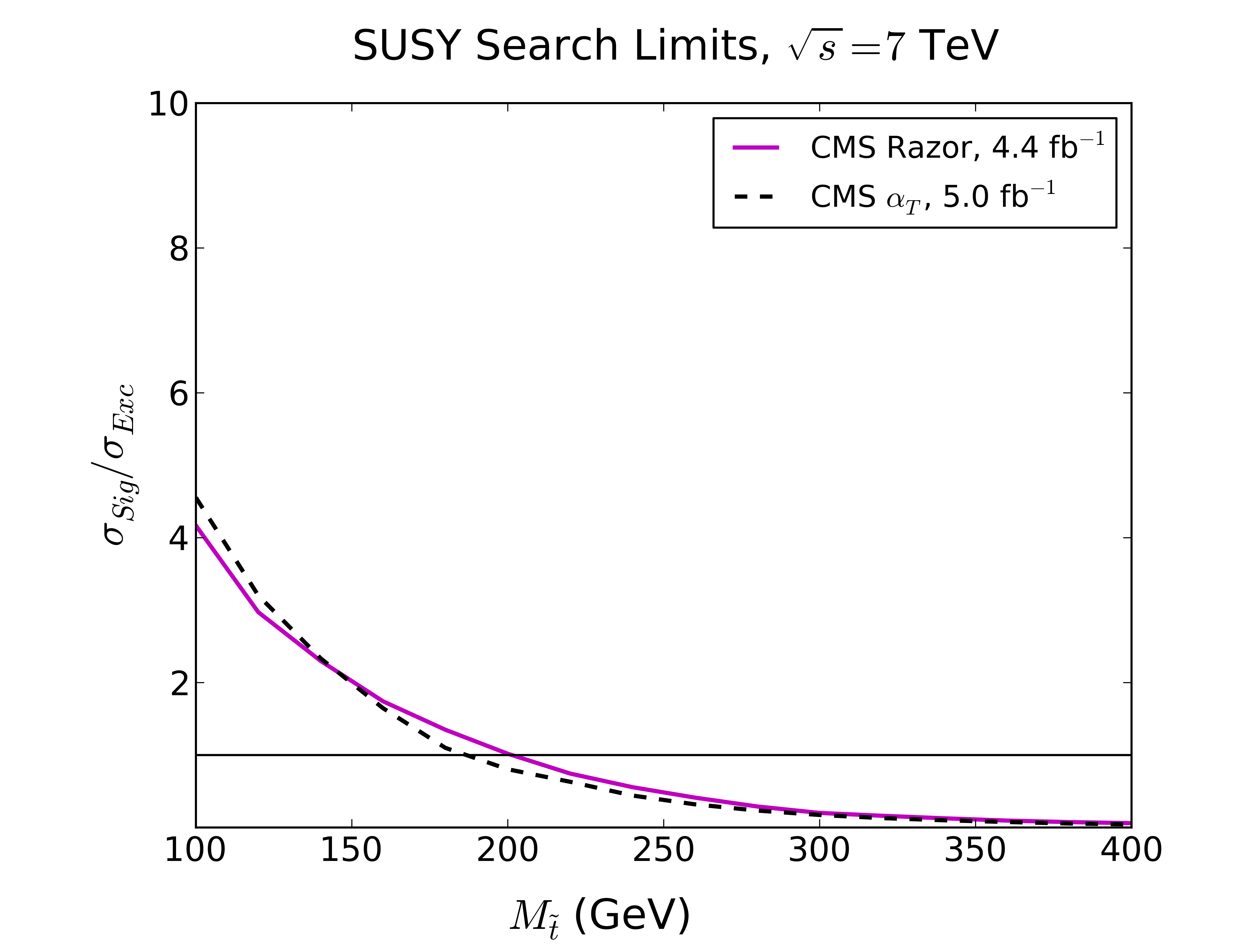}
  \caption{Limits from the monojet and SUSY searches for stops (or a
  single eigenstate squark) in the limit of quasi-degeneracy with the
  LSP.\label{fig:StopLineLimit}}
\end{figure*}

\begin{figure*}
  \centering
  \includegraphics[width=0.49\textwidth]{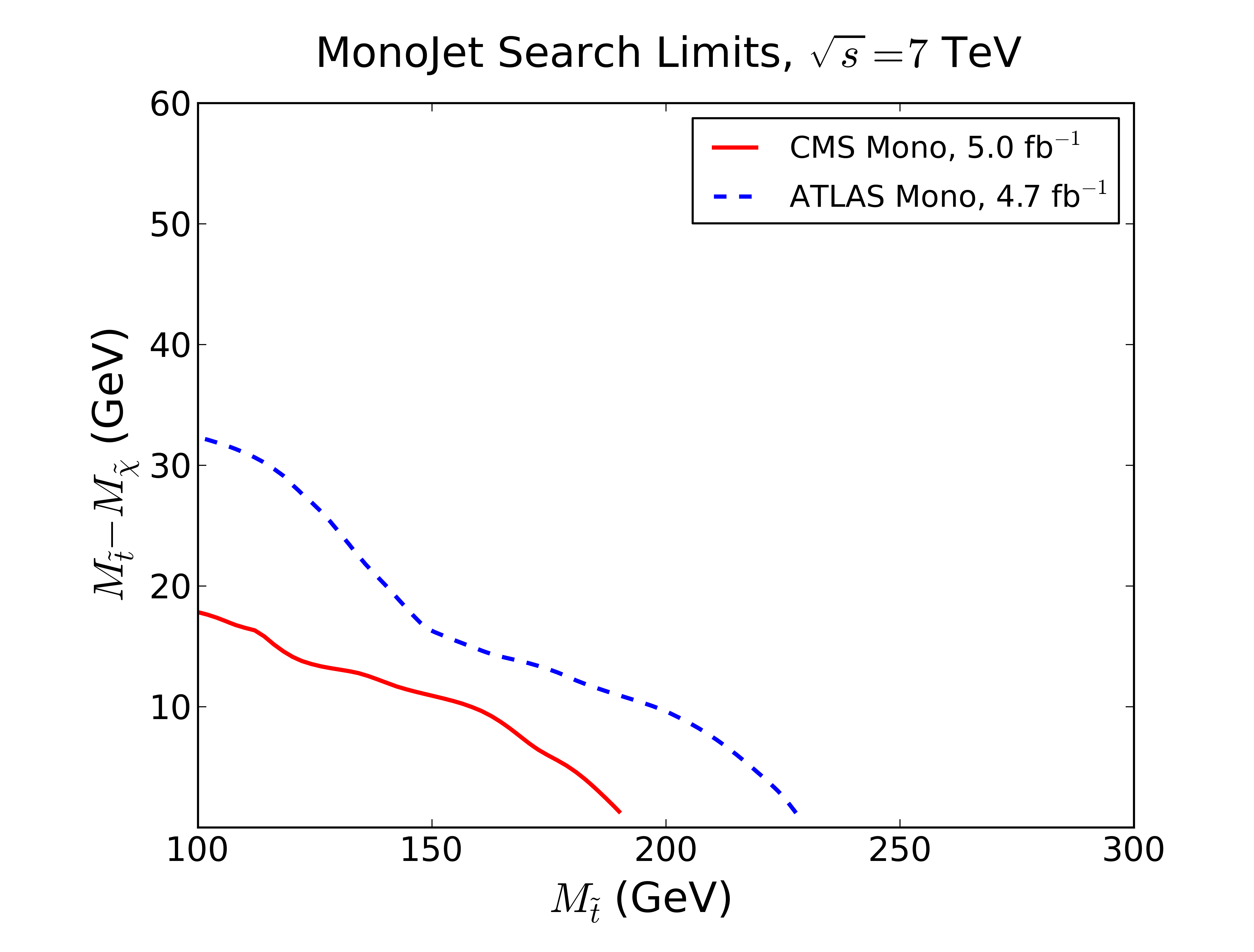}
  \includegraphics[width=0.49\textwidth]{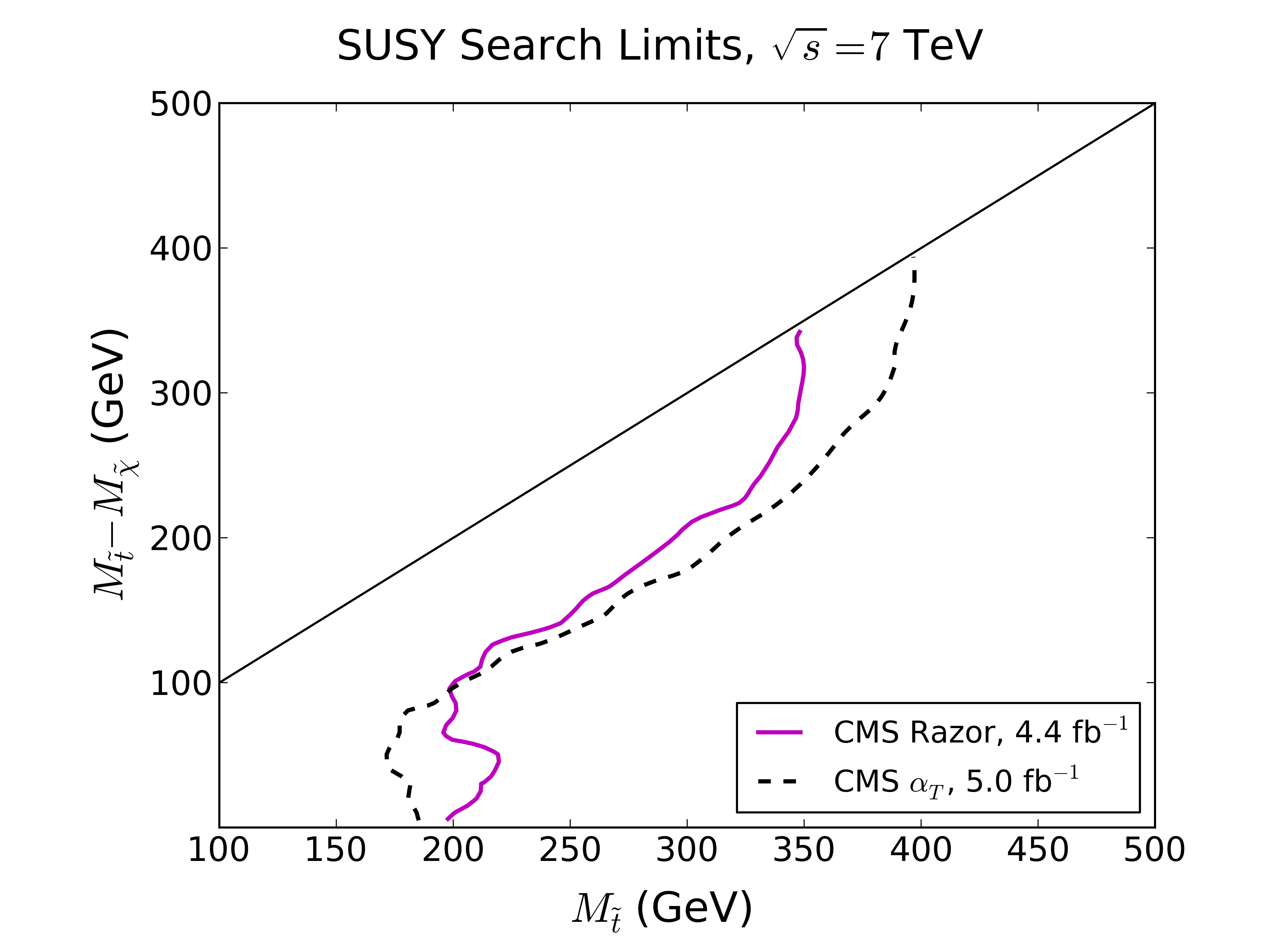}
  \caption{Limits from the monojet and SUSY searches for stops (or a
  single eigenstate squark) as the mass splitting  to the LSP
  is increased.\label{fig:StopLoopLimit}}
\end{figure*}

To calculate the limits on each of our degenerate simplified models we
use the latest NLO+NLL SUSY cross-sections from NLL-Fast
\cite{Beenakker:1996ch,Beenakker:2009ha,Beenakker:2011fu}. The
theoretical uncertainty is calculated including the factorisation and
renormalisation scale and parton density functions (PDF)
\cite{Nadolsky:2008zw} errors for both the matched distributions and
the total cross-section. In addition, we vary the different scales involved in the matching algorithms and
the parton showers, \textit{cf.} 
Fig.\ref{fig:MGM1vsP8M2}, and take the result with the least
constraining bound. In the limit of mass degeneracy with the LSP, all
bounds from the different searches are given in
Tab.~\ref{tab:Limits}.

\subsection{Stop (single eigenstate) Limit}

We begin by describing the limits that apply to either stops in
compressed spectra or more generally a single eigenstate squark,
with the gluino and all other squarks very heavy. In the limit of degeneracy
we find that the ATLAS monojet search provides the best limit of
$m_{\tilde{t}_1}>230$~GeV, Fig~\ref{fig:StopLineLimit}. In contrast
the CMS monojet search is less constraining with a limit of
$m_{\tilde{t}_1}>190$~GeV.

\begin{figure*}
  \centering
  \includegraphics[width=0.49\textwidth]{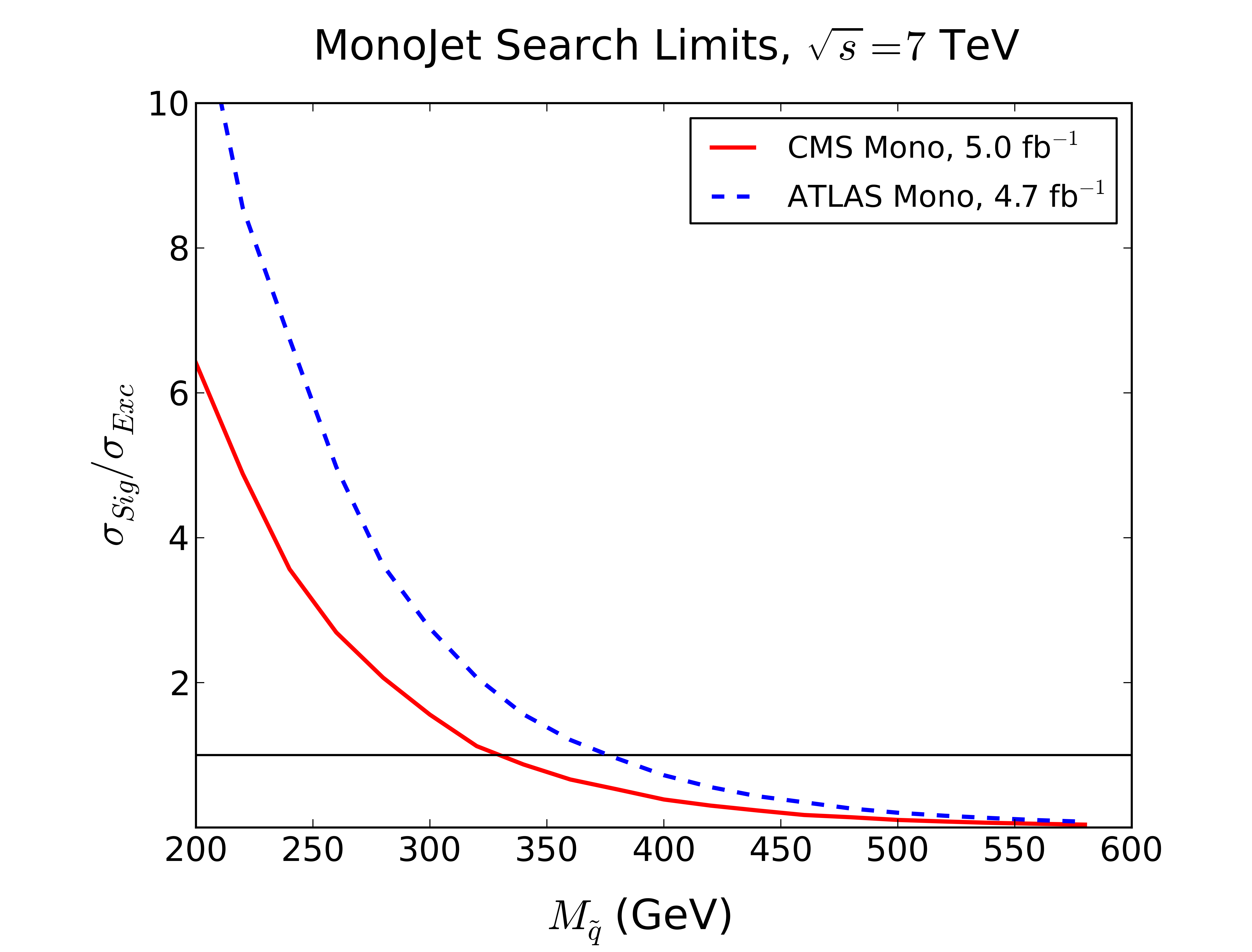}
  \includegraphics[width=0.49\textwidth]{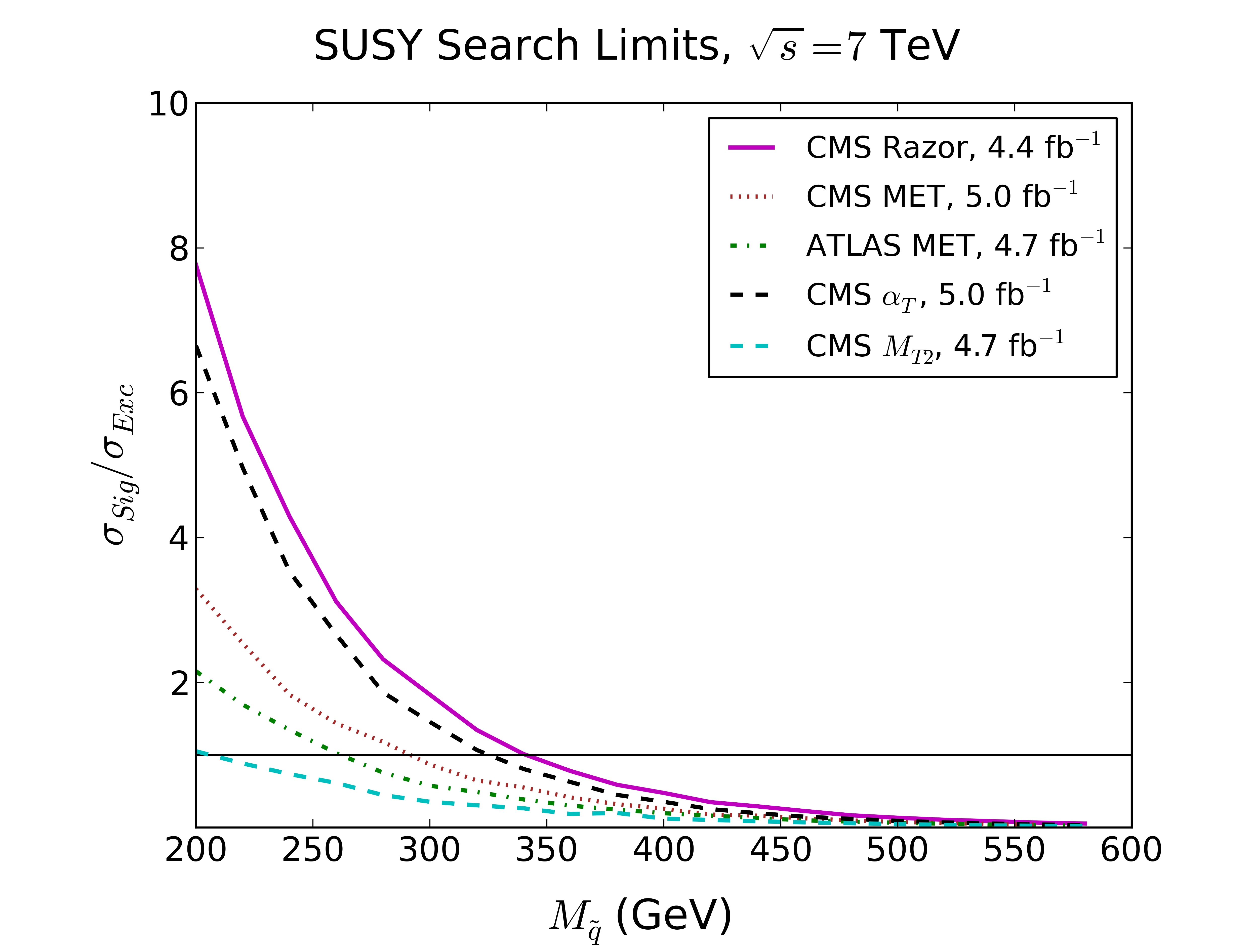}
  \caption{Limits from the monojet and SUSY searches for first and
  second generation squarks in the limit of quasi-degeneracy with the
  LSP.\label{fig:SquarkLineLimit}}
\end{figure*}

\begin{figure*}
  \centering
  \includegraphics[width=0.49\textwidth]{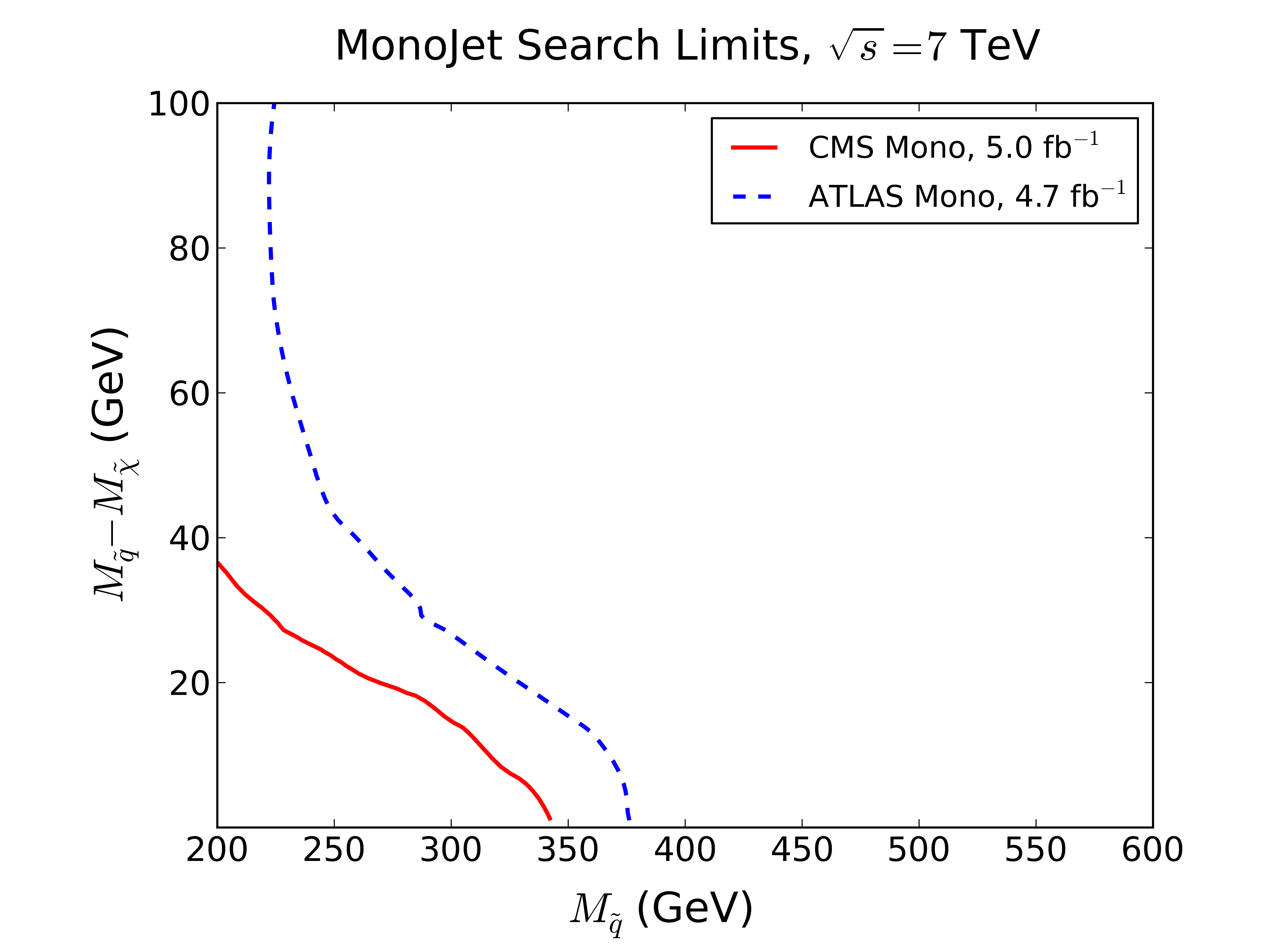}
  \includegraphics[width=0.49\textwidth]{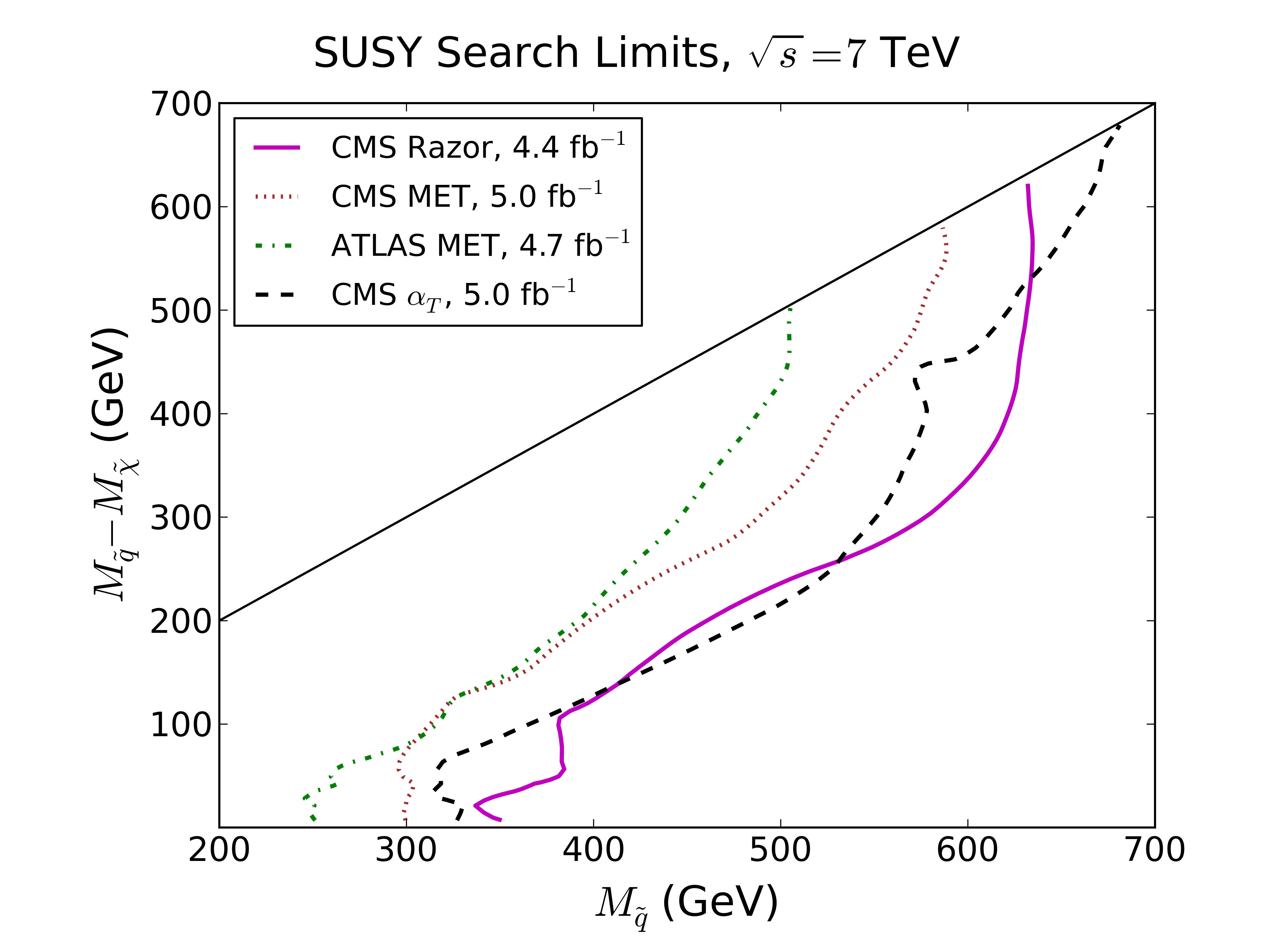}
  \caption{Limits from the monojet and SUSY searches for first and
  second generation squarks as the mass splitting to the LSP is
  increased.\label{fig:SquarkLoopLimit}}
\end{figure*}

The reason that the ATLAS search provides a better limit despite the
signal selection for the two searches being very similar is a
reduction on the error of the main background, $Z \to \nu \bar{\nu} +
jets$. To estimate this background, both experiments use a data driven
technique that measures other visible electroweak processes in control
regions and only uses transfer functions from the Monte Carlo
program to find the backgrounds. In the CMS analysis, only the
process $Z \to \mu^- \mu^+ + jets$ is used as a control region and
this is done by removing the muons and treating them as missing energy
vectors. However, the limiting factor in this analysis is the sample
size of $Z \to \mu^-\mu^+ + jets$ and the resulting
statistical error. The ATLAS analysis partially
remedies this situation by also using $Z\to e^-e^+ + jets$,
$W\to\mu\nu+jets$ and $W\to e\nu+jets$ to increase the number of
events. These backgrounds have larger associated systematics but this
problem is outweighed by the reduction in the total statistical error.

We also find that some of the general SUSY searches can be competitive
even in the limit of degeneracy. The CMS razor search provides a limit
of $M_{\tilde{t}_1}>200$~GeV, while CMS $\alpha_T$ gives a
limit of $M_{\tilde{t}_1}>190$~GeV. Unfortunately the more traditional `Jet and MET' searches had too small efficiencies
to provide reliable limits for this particular simplified model. A
major difference is the lower kinematic acceptances allowed by the
topology based searches that give access to smaller SUSY mass states.

As the mass splitting, $M_{\tilde{t}_1} - M_{LSP}$, is increased,
Fig.~\ref{fig:StopLoopLimit}, we find that the monojet searches
rapidly lose their effectiveness. In fact, as soon as $M_{\tilde{t}_1}
- M_{LSP} > 30$~GeV, we can no longer set a reliable limit with these
analyses. The reason is that both monojet searches include a third jet
veto, $p_T>30$~GeV. Thus, any extra radiation in the final state
produced by SUSY decays will increase the likelihood of a third jet
being present and these events will be vetoed.

In contrast, the SUSY searches are stable as the mass splitting is
increased to $M_{\tilde{t}_1} - M_{LSP} = 100$~GeV. In this region we
have a balance between extra radiation from decays increasing the
number and hardness of final state jets against a reduction in the
momentum carried by the LSP and thus less missing energy. However,
once the mass splitting goes beyond this point, $M_{\tilde{t}_1}-M_
{LSP} > 100$~GeV, the extra jet activity rapidly increases the limit we obtain. For a massless LSP, the
CMS $\alpha_T$ search provides the best limit in this model of
$M_{\tilde{t}_1}>400$~GeV.

\begin{figure*}
  \centering
   \includegraphics[width=0.49\textwidth]{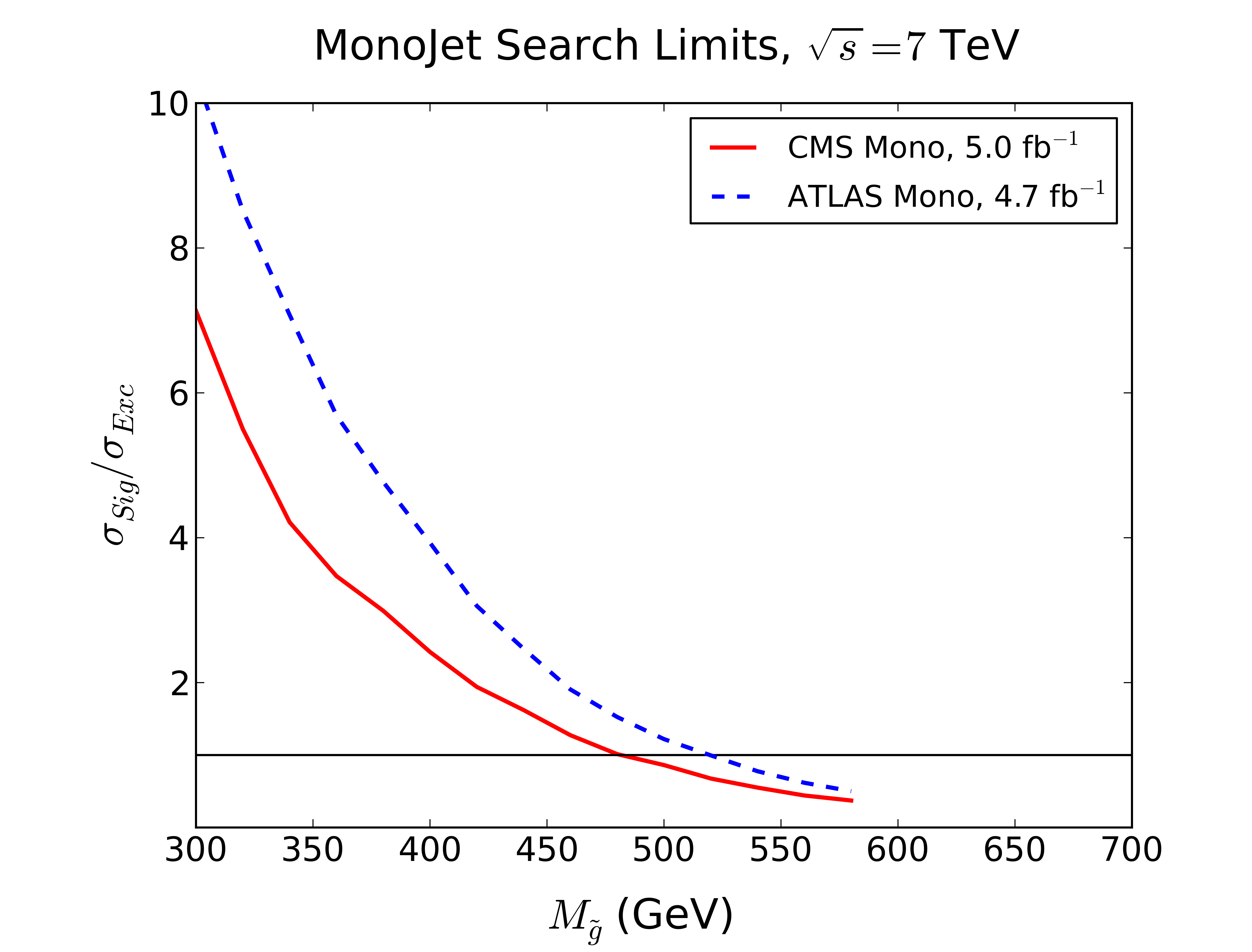}
   \includegraphics[width=0.49\textwidth]{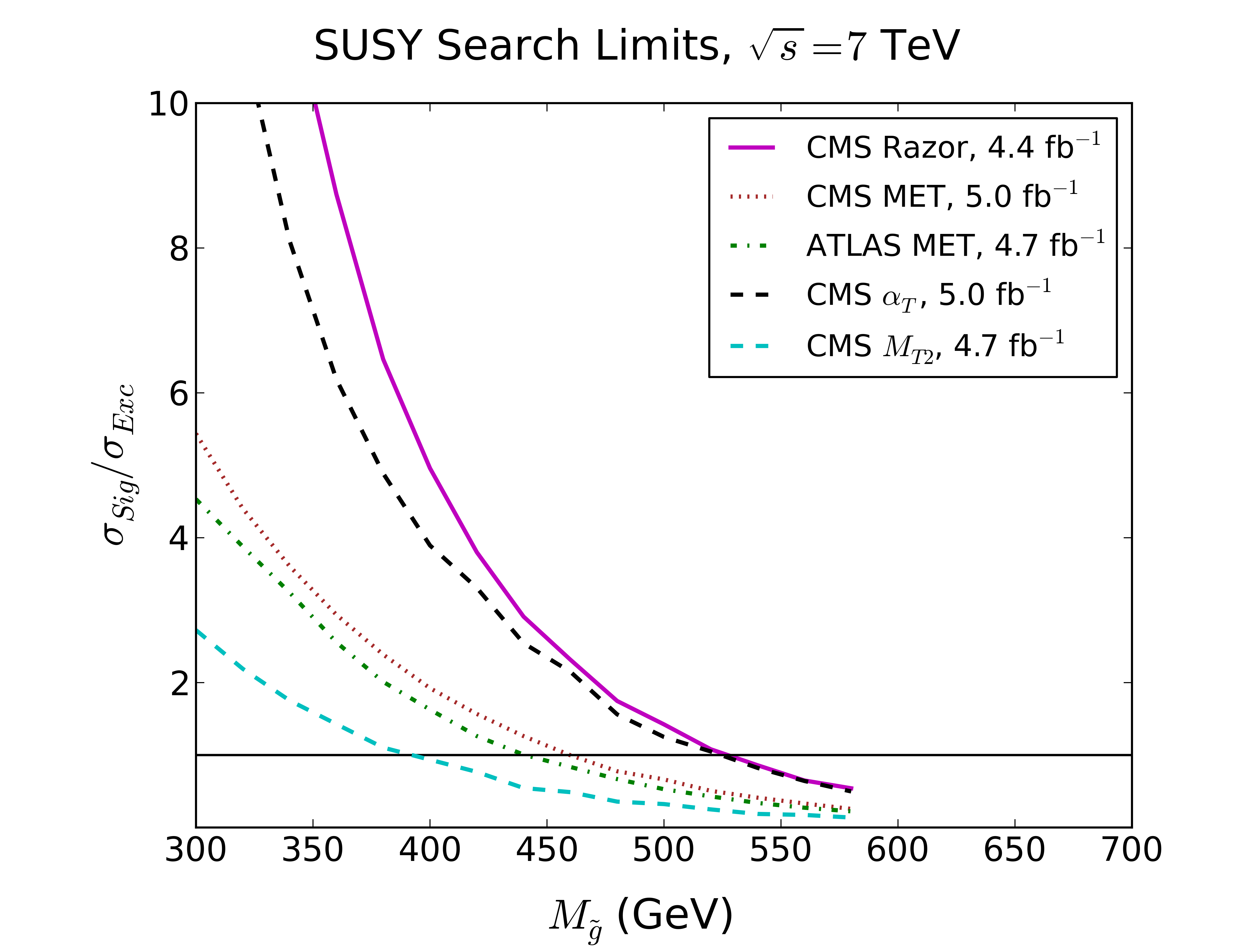}
   \caption{Limits from the monojet and SUSY searches for gluinos in the limit of quasi-degeneracy with the LSP.\label{fig:GluLineLimit}}
\end{figure*}

 \begin{figure*}
   \centering
   \includegraphics[width=0.49\textwidth]{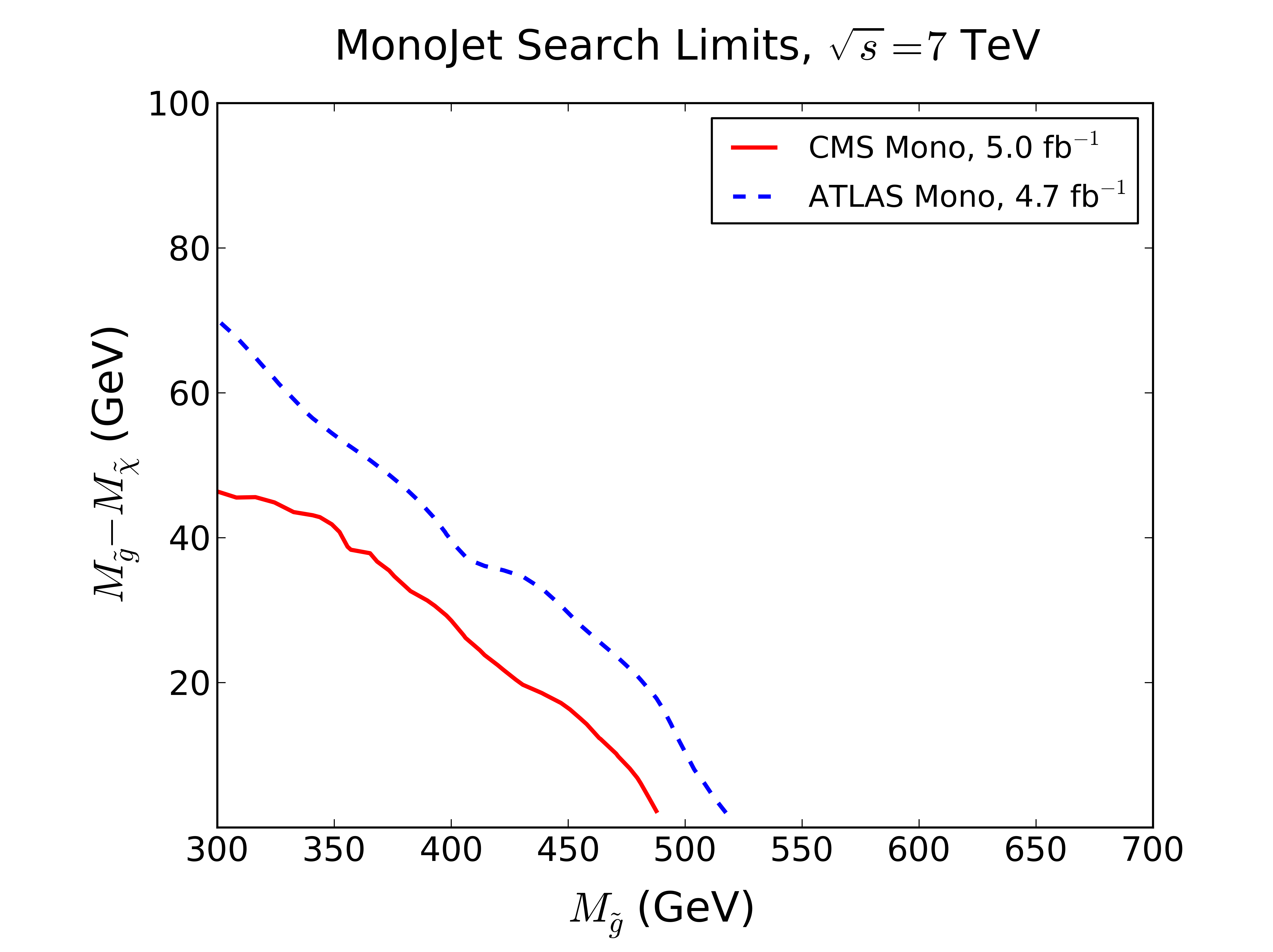}
   \includegraphics[width=0.49\textwidth]{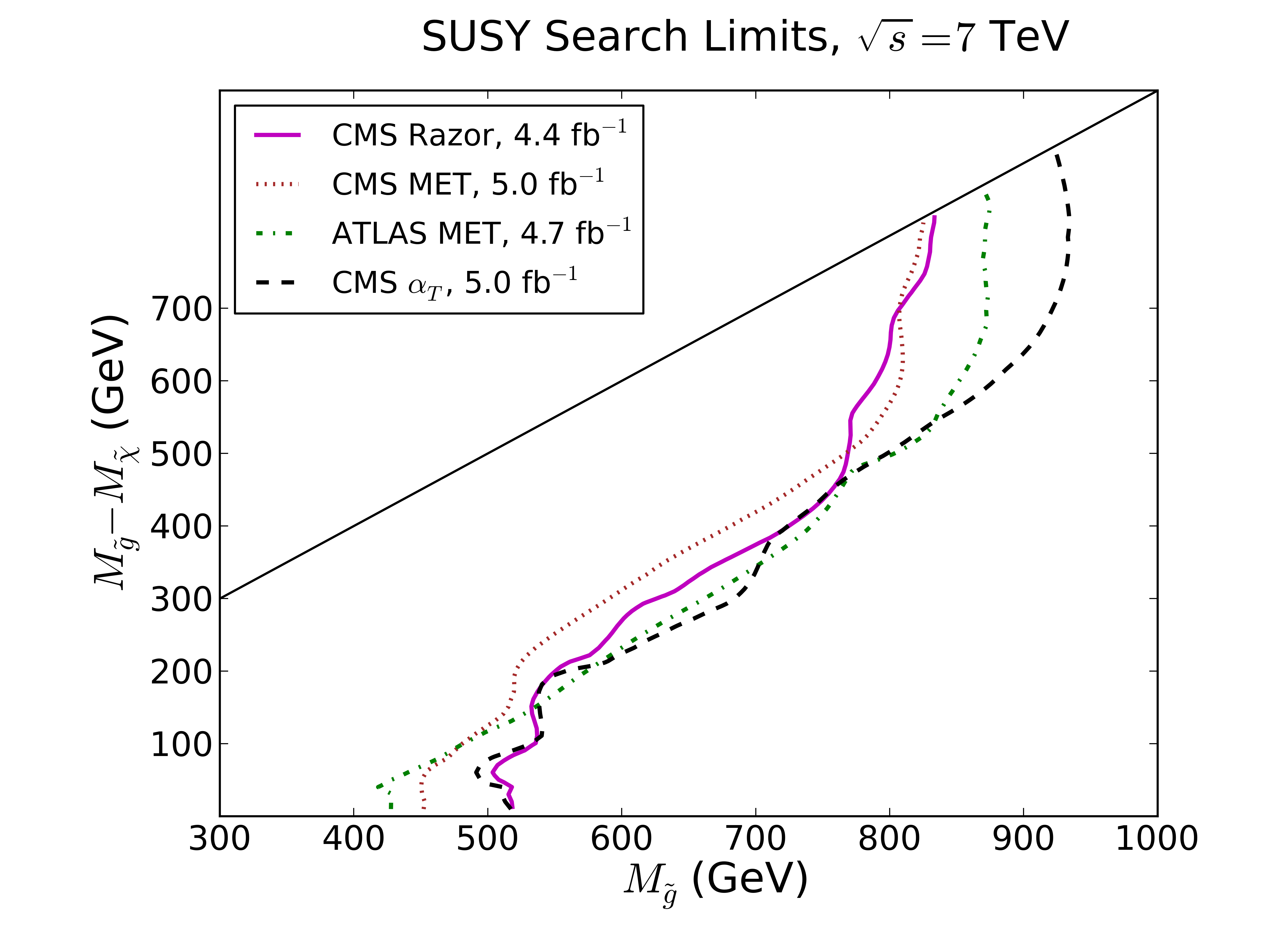}
   \caption{Limits from the monojet and SUSY searches for gluinos as the mass splitting to the LSP is increased. \label{fig:GluLoopLimit}}
 \end{figure*}

We must however remind the reader that the limits for increased mass
splitting are only applicable in the case of the decay $\tilde{t}_1
\to LSP + j$. For stops, once the mass splitting is increased to
$M_{\tilde{t}_1} - M_{LSP} > M_{t}$ this is very unlikely to be the
case in SUSY models. In this region it is only sensible to consider
this as a limit on a single eigenstate squark from the first or second
generation. Even in these models however, this assumes a single step
decay and more complicated topologies may reduce the bound.

\subsection{Squark Limit}

We next explore the limits for a simplified model where the first and
second generation squarks are degenerate with the LSP,
Fig.~\ref{fig:Spectrum}(b). In the limit of degeneracy, the ATLAS
monojet search places the tightest bound on the scenario of
$M_{\tilde{q}}>370$~GeV, Fig.~\ref{fig:SquarkLineLimit}. The CMS
monojet bound is again a little weaker ($M_{\tilde{q}}>340$~GeV) due
to the larger error quoted on the $Z \to \nu \bar{\nu} + jets$
background.

In this case, the topology based SUSY searches are competitive
and again the CMS razor search gives the best limit of
$M_{\tilde{q}}>350$~GeV. Due to the increased mass scale of the
produced particles, the normal `Jets and MET' searches now have a
large enough acceptance efficiency to provide reliable
limits. However, these are significantly lower at
$M_{\tilde{q}}>300$~GeV for CMS and $M_{\tilde{q}}>260$~GeV for
ATLAS. We notice that the search regions that give the most
discriminating search power are those with the highest proportion of
missing energy in the event and relatively softer cuts on jets,
Tab.~\ref{tab:Limits}.

As the mass splitting between the squarks and the LSP ($M_{\tilde{q}}
- M_{LSP}$) is increased, we again see the same trend as for the
simplified stop model searches, Fig.~\ref{fig:SquarkLoopLimit}. The
monojet searches immediately lose their power and
when the mass splitting is only 40~GeV, the ATLAS monojet bound has
been reduced to just $M_{\tilde{q}}>250$~GeV (the CMS search can no
longer place any bound).

\begin{figure*}
  \centering
  \includegraphics[width=0.49\textwidth]{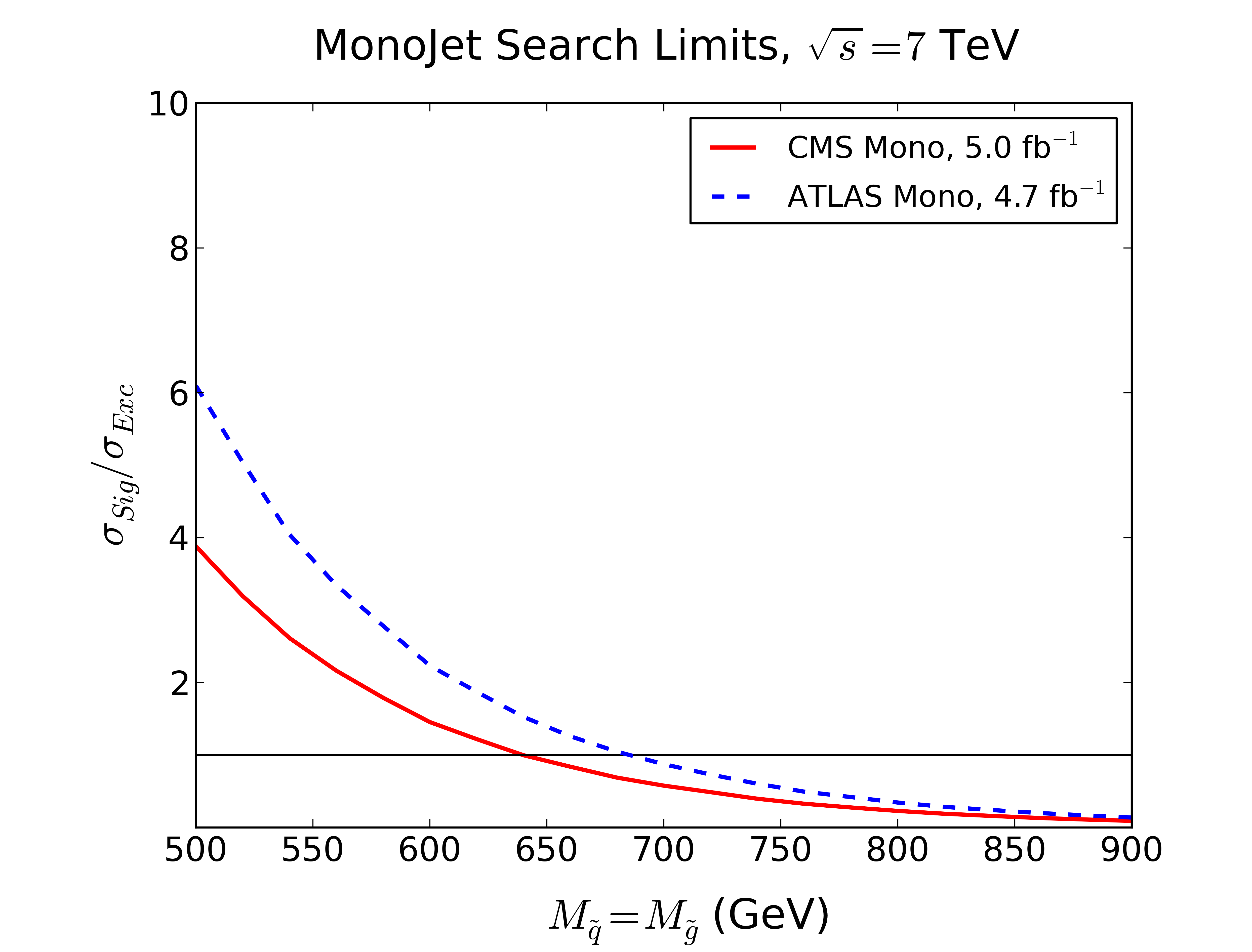}
  \includegraphics[width=0.49\textwidth]{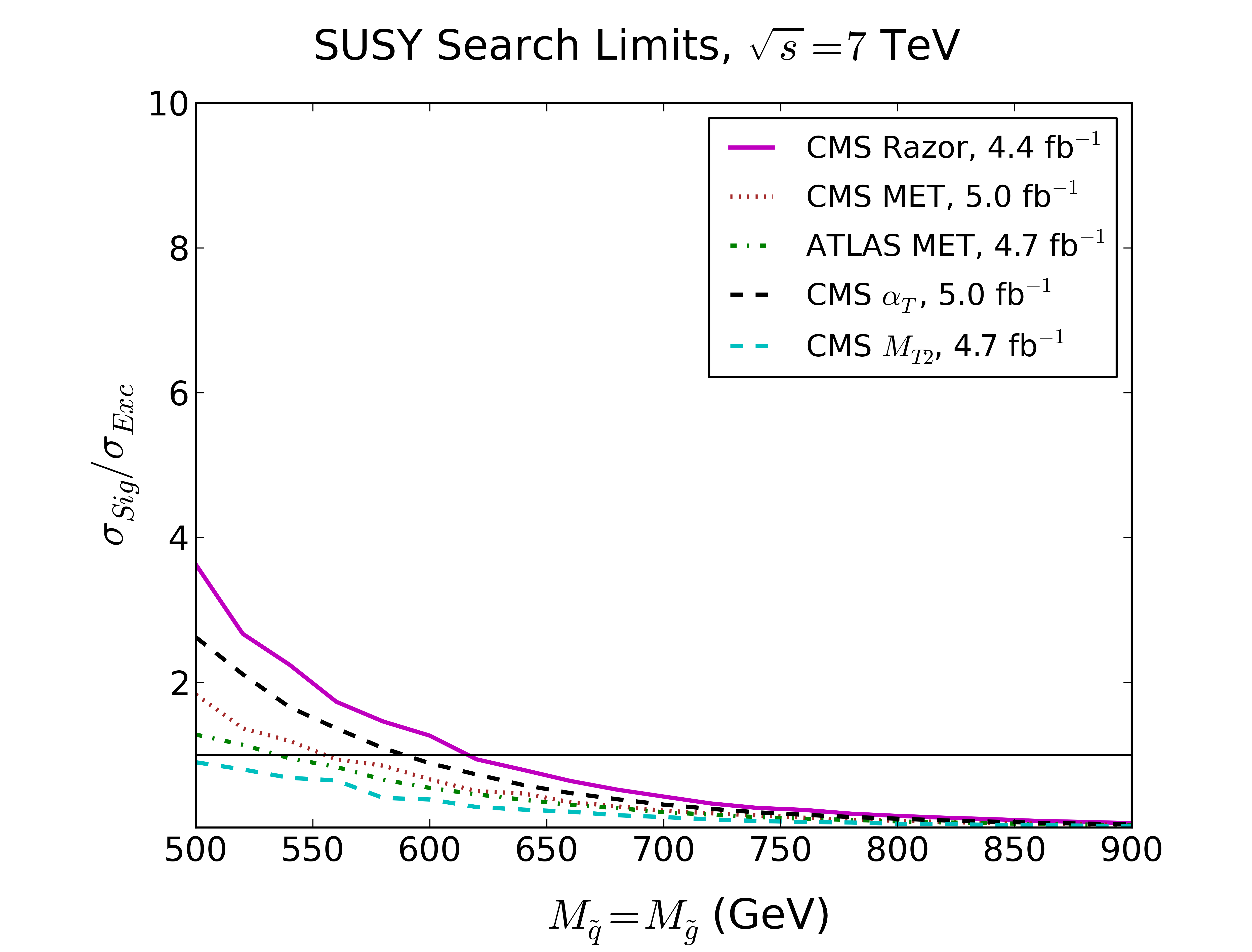}
  \caption{Limits from the monojet and SUSY searches for the equal
  mass squark, gluino scenario in the limit of quasi-degeneracy with
  the LSP. \label{fig:TotLineLimit}}
\end{figure*}

\begin{figure*}
  \centering
  \includegraphics[width=0.49\textwidth]{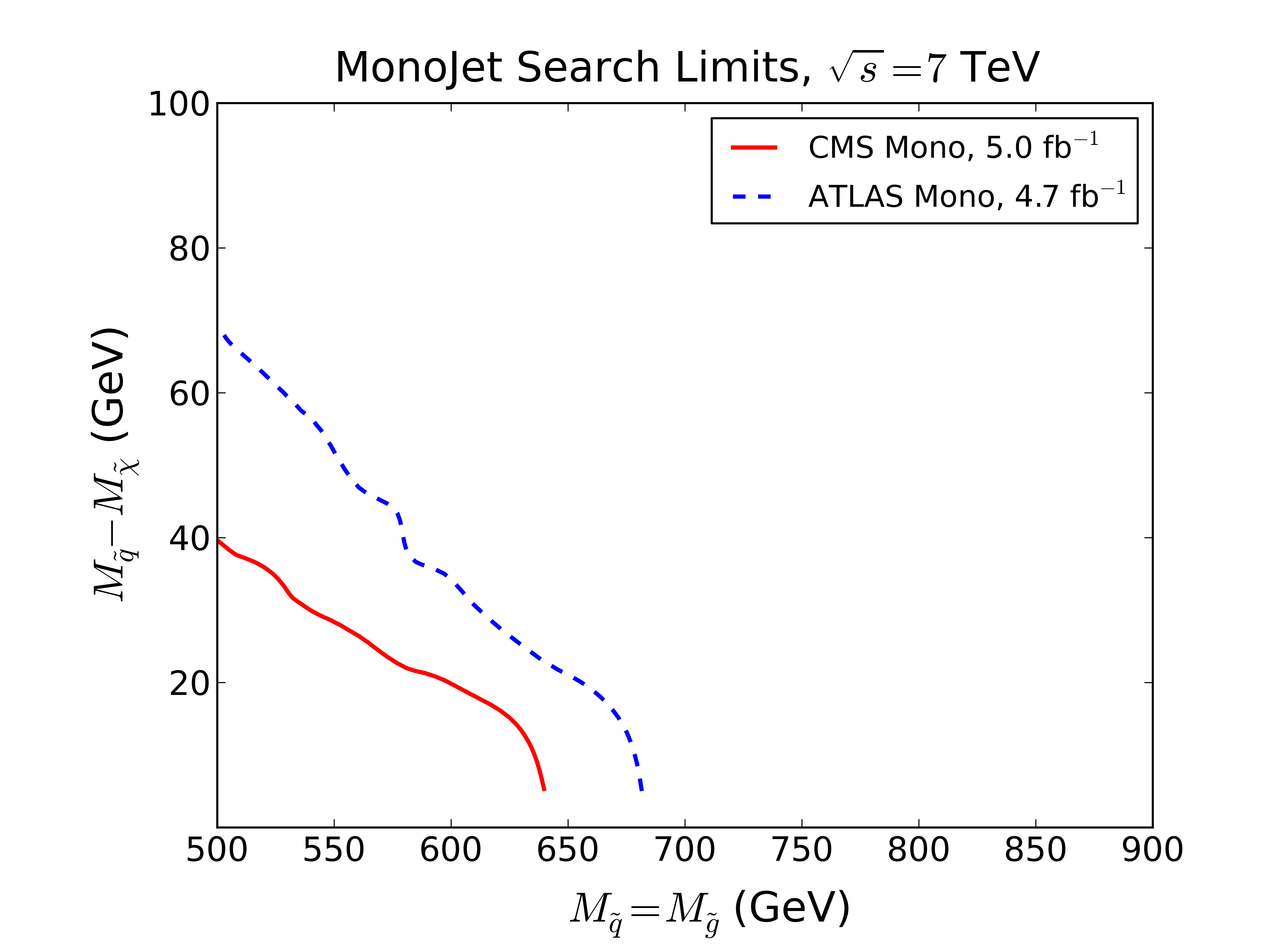}
  \includegraphics[width=0.49\textwidth]{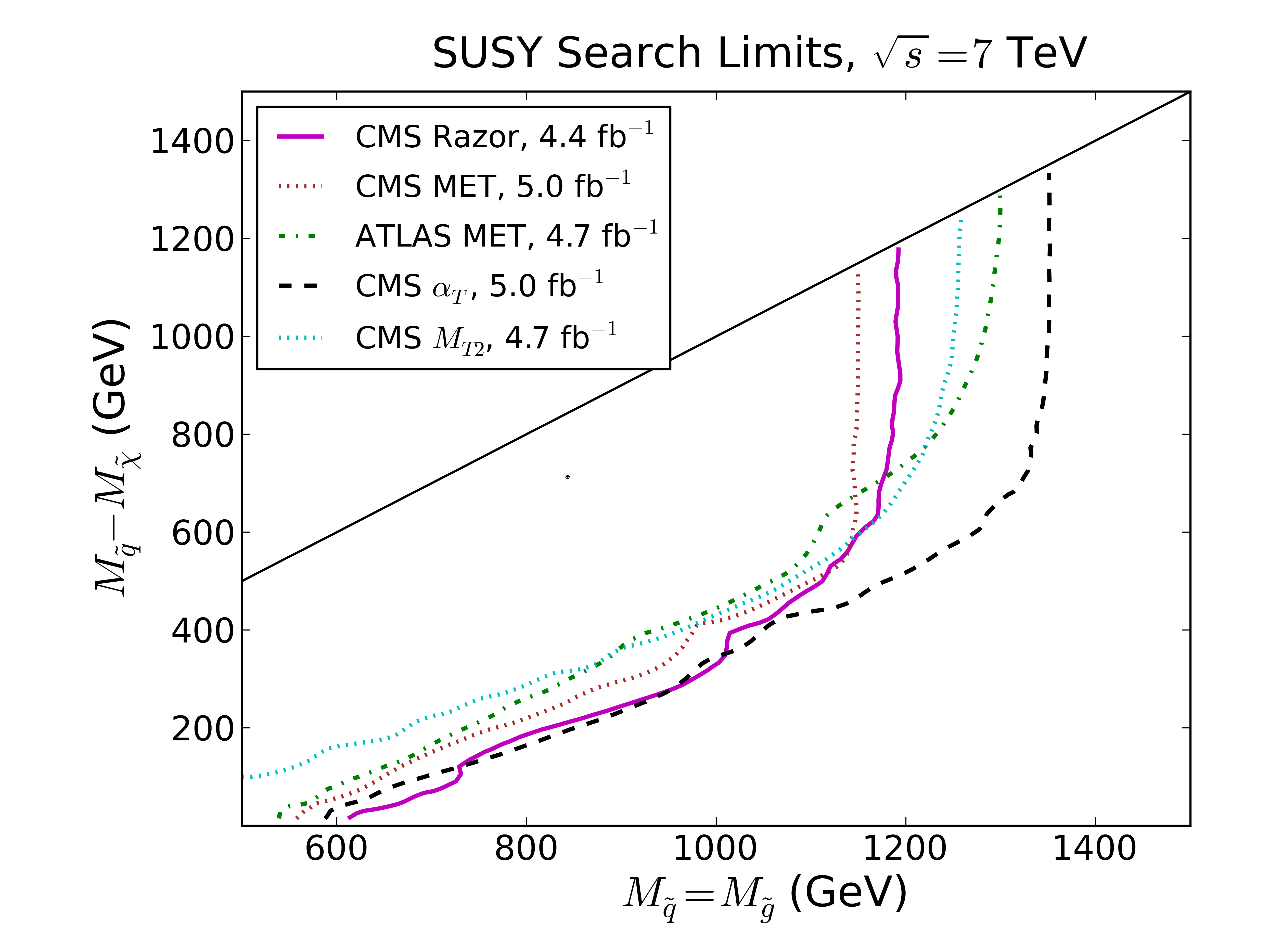}
  \caption{Limits from the monojet and SUSY searches for the equal
  mass squark, gluino scenario as the mass splitting to the LSP is
  increased. \label{fig:TotLoopLimit}}
\end{figure*}

However, the SUSY searches are again stable as the mass splitting is
increased and actually improve more rapidly than for the stop
searches. This is due to the limits being in a different kinematic
regime, where the reduction in missing momentum due to increased mass
splitting occurs more slowly.

One may notice that the SUSY limits are not smooth as the mass
splitting is increased but instead often show discontinuities across
the parameter space. The source of these discontinuities comes from
the fact that we set limits only using the single search region that
produces the most constraining bound. As we move across the parameter
space we jump between different search regions and the discontinuities
lie at these intersections. If we instead set limits by combining all
search regions into a single variable, these would be removed and a
more constraining bound may be produced.

\subsection{Gluino Limit}

For the simplified gluino model we find that the SUSY based, CMS razor
and $\alpha_T$ ($M_{\tilde{g}}>530$~GeV) now set the best limits,
Fig.~\ref{fig:GluLineLimit}. However the ATLAS monojet search is still
very competitive with a bound of $M_{\tilde{g}}>520$~GeV in the limit
of degeneracy with the LSP.

\begin{figure*}
  \centering
  \includegraphics[width=0.65\textwidth]{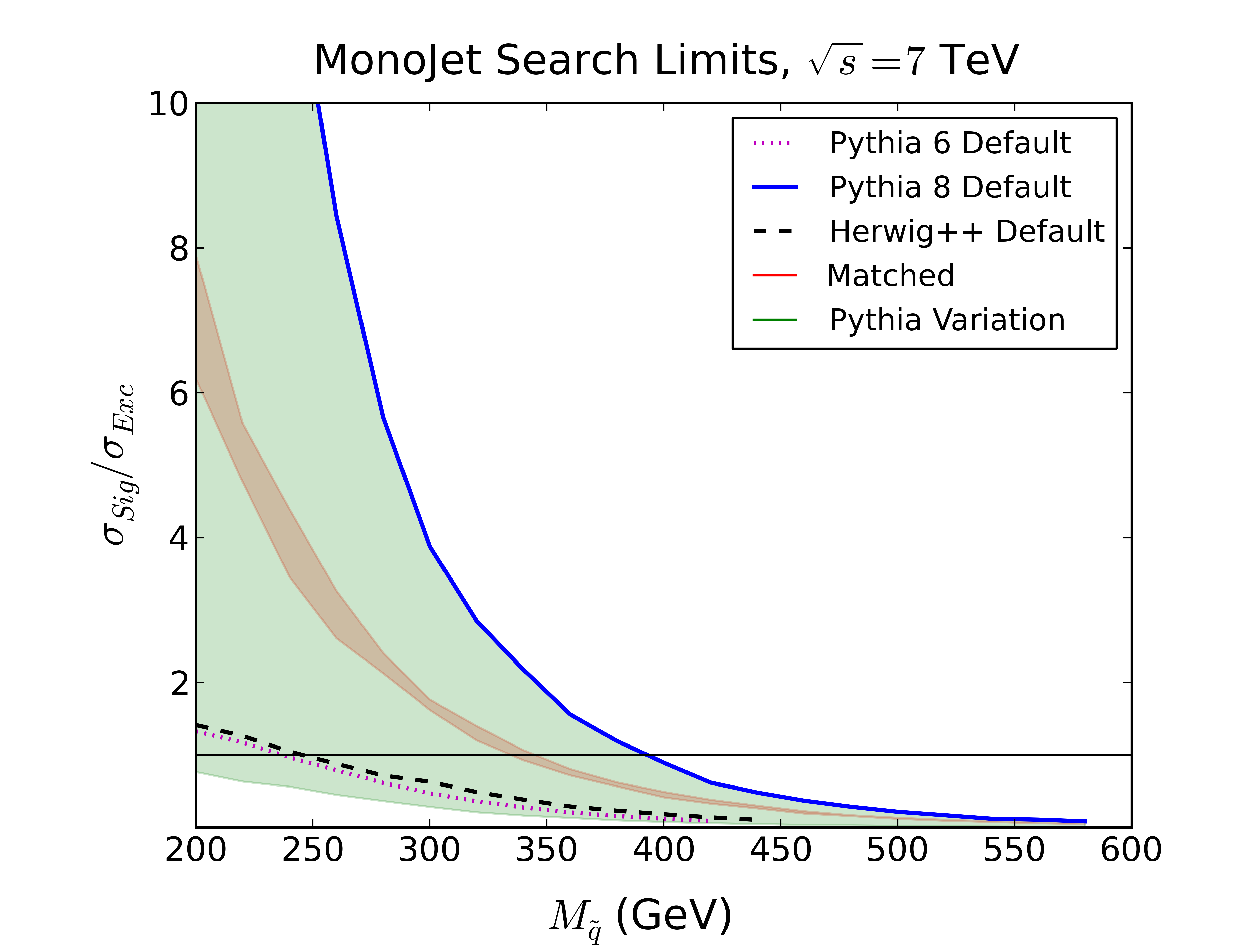}
  \caption{Comparison of the limit found when using matching with that
  when only using different parton shower choices. The light green
  area shows the variation in the limit when the Pythia 6 parton
  shower is varied between 'wimpy' and 'power' settings. The red area
  shows the variation in the limit as the matching scale is varied
  between 50 and 200~GeV, the factorsation and renormalisation scales
  are varied between $M_T/2$ and $2M_T$ and the parton shower between
  `wimpy' and `power' settings. \label{fig:PSLineLimit}}
\end{figure*}

Once again, as the mass splitting is increased the monojet searches
lose all of their power while the SUSY searches 
remain stable, Fig.~\ref{fig:GluLoopLimit}. However, there are some
important differences in how the limits evolve with the mass
splitting compared to previous models. Firstly, we now see that the
limits from the SUSY searches do not begin to rapidly improve until
the mass splitting, $M_{\tilde{g}} - M_{LSP} > 200$~GeV. This occurs
due to the three body decays of the gluino that share the energy of
the decay between two final state jets instead of just
one. Consequently, energy is less focused into single hard jets and is
more likely to produce a multi-jet signal that is harder to separate
from background.

\subsection{Equal Mass Limit}

The fourth simplified model we consider has the
first and second generation squarks and the gluino all degenerate with
the LSP. We find that in the limit of degeneracy, the most
constraining limit is given by the ATLAS monojet search with
$M_{\tilde{q}} \sim M_{\tilde{g}} >680$~GeV,
Fig.~\ref{fig:TotLineLimit}. The SUSY searches also provide
competitive limits with CMS razor being the most constraining
($M_{\tilde{q}} \sim M_{\tilde{g}} >610$~GeV).

As we have seen for the previous models, the monojet searches again
fail as the mass splitting is increased from vetoed events due to
extra jets. For example, once the mass splitting $(M_{\tilde{q}} \sim
M_{\tilde{g}}) - M_{LSP} > 70$~GeV, the limit from the ATLAS monojet
search falls below 500~GeV. In contrast, the limits from SUSY searches
increase as soon as the mass splitting is increased and for  CMS
$\alpha_T$ reach 1400~GeV for a massless LSP.

An interesting feature of the evolution of the limits with increased
mass splitting is how the ATLAS MET search provides a very competitive
limit of 1300~GeV for a low mass LSP\footnote{This is slightly lower
than the limit presented by the official ATLAS analysis due to the
more conservative limit setting procedure we use here, see
Sec.~\ref{sec:Searches}.}\!. We believe that this is due to the fact
that this search has been primarily designed with the popular CMSSM in
mind. In the CMSSM, large mass splittings between the coloured
particles and the LSP are always present and thus the search is tuned
for these topologies. However, as we have shown, for more compressed
topologies, more general search strategies can lead to a more powerful
result.

\subsection{Parton Shower Comparison}

Here we discuss how the results depend on the method
chosen to simulate ISR in our simplified models. We consider
the CMS monojet search for first and second generation squarks
degenerate with the LSP as an example. We compare our matched
prediction with those coming from the various parton shower choices
that can be made, Fig.~\ref{fig:PSLineLimit}. One can see that the
uncertainty in the limit associated with the theoretical error on the
matching prediction is $\pm 5$~GeV, with a lower limit of
$M_{\tilde{q}} > 340$~GeV. In contrast, the range in limits that
different parton shower choices and settings gives is $\pm 110$~GeV,
with a lower limit $M_{\tilde{q}} > 180$~GeV for the softest jet
distributions and up to $M_{\tilde{q}} > 400$~GeV for the hardest.

Even if we only take the limits given by the default parton shower
behaviour, we still see a large range in the predictions. For example
Herwig++ and Pythia 6 both give a limit in this scenario of
$M_{\tilde{q}} > 250$~GeV, while Pythia 8 with its power shower as
default leads to a limit of $M_{\tilde{q}} > 400$~GeV. Once again we
would like to point out that the default Pythia 8 settings can be in
contradiction with the naive picture of parton showers always
producing radiation that is softer than the matrix element.

\subsection{Possible Search Improvements}

The complementarity of the SUSY searches with the monojet analyses to
our simplified models motivates the question of whether the search
strategy can be improved. We do not quantitavily address this question
here but instead offer some suggestions that will be investigated in
future.

We have seen that in the limit of degeneracy, the monojet and
topological CMS SUSY searches (razor and $\alpha_T$) produce very
similar limits. However, the searches are (somewhat) orthogonal
because the SUSY searches require two hard jets (razor:
$p_T(j)>50$~GeV, $\alpha_T$: $E_T(j)>100$~GeV) whilst the monojet
searches veto events with a third jet ($p_T(j_3)>30$~GeV). This leads
to the obvious question of whether either of these restrictions could
be relaxed to improve the search reach.

In the case of the SUSY searches we see no reason why an extra search
region could not be included that also allows a monojet topology as
well as the `normal' multijet toplogies. Whilst the $Z+jets$
background may be enhanced in this region, we feel that such a search
would offer an improvement for very compressed spectra.

For the monojet toplogies, we would be interested in a search that
removed the third jet veto requirement in order not to remove signal
events with additional soft jets. For example, such a veto could be
replaced by a geometrical cut that requires all jets to lie within an
azimuthal region of size, $\phi=2.5$, for example. Another
idea would be to instead search for a far larger monojet, perhaps
$R=2.5$, as this would sum all radiation together. Both these
approaches should not increase the QCD background since back to back
jets will still not be accepted. We hope to study both these proposals
in more detail soon.

 \section{Conclusion} {\label{sec:Conclusion}}
 We have considered the case where the supersymmetric spectrum is highly compressed. This invalidates most of the previous searches at the LHC, since the decay products are very soft and do not pass the experimental cuts. We have instead proposed to use hard initial state QCD radiation to set limits on various compressed simplified SUSY models. The models considered the possibility that squarks and/or gluinos could be degenerate with the LSP. We then compared various existing ATLAS and CMS LHC searches and set lower mass limits in our simplified scenarios. In addition, the breaking of the mass degeneracy was investigated and we show how the bounds evolve, as we increase the mass splitting.

Vital to the study was a reliable prediction of the hard QCD radiation. We matched the matrix element to the parton shower in order to calculate as precise a prediction as possible. Two matching algorithms were investigated and different scales and parton shower settings were chosen to find an estimate of the associated uncertainty.

To set bounds we used all the current 7~TeV hadronic SUSY searches as well as the current monojet searches. We found that in the limit of degeneracy, the topological SUSY searches (razor and $\alpha_T$) and monojet searches gave the most constraining bounds. However, as soon as the degeneracy is broken, the monojet searches quickly lose their effectiveness. In contrast, the SUSY searches are stable and once the mass splitting increases above $\sim$ 100~GeV, the bounds rapidly improves.

For the `Stop (single eigenstate)' scenario we found a bound $M_{\tilde{t}_1}>230$~GeV. By considering the first two generations of squarks, (`Squark' scenario) we essentially increase the cross-section of production and this leads to a bound of $M_{\tilde{q}}>370$~GeV. In the next scenario, we completely decouple the squarks and place the gluino degenerate with the LSP (`Gluino' scenario) which gives a bound, $M_{\tilde{g}}>530$~GeV. As a fourth scenario, we take both the gluino and squark degenerate with the LSP (`Equal mass' scenario) and this places a limit of $M_{\tilde{q}}\sim M_{\tilde{g}} > 680$~GeV.

\acknowledgments

\noindent We would especially like to thank Stefan Prestel for his help adapting the Pythia 8 matching algorithm. In addition we 
would like to acknowledge useful discussions with John Conley, Krysztof Rolbiecki, Daniel Wiesler, Johan Alwall and Maurizio 
Pierini. This work has been supported in part by the Helmholtz Alliance `Physics at the Terascale' and the DFG SFB/TR9 ``Computational Particle
Physics''. HKD was supported by BMBF Verbundprojekt HEP-Theorie� under the contract 0509PDE.


\bibliographystyle{utphys}
\bibliography{ISR_SUSY}

\end{document}